\documentclass[twocolumn]{aastex631}

\shortauthors{Davis et al.}
\graphicspath{{./}{figures/}}

\newcommand{\difd}{\mathrm{d}}

\usepackage{amsmath}

\newcommand{\tleco}{\texttt{Tleco}}

\begin{document}

\title{Tleco: A Toolkit for Modeling Radiative Signatures from Relativistic Outflows}

\email{zkd@purdue.edu.edu}

\author[0000-0002-0959-9991]{Zachary Davis}
\affiliation{Department of Physics, Purdue University, 525 Northwestern Avenue, West Lafayette, IN 47907, USA}

\author[0000-0003-1988-1912]{Jesús M. Rueda-Becerril}
\affiliation{Department of Physics, Purdue University, 525 Northwestern Avenue, West Lafayette, IN 47907, USA}

\author[0000-0003-1503-2446]{Dimitrios Giannios}
\affiliation{Department of Physics, Purdue University, 525 Northwestern Avenue, West Lafayette, IN 47907, USA}


%
%
%
%
%



\begin{abstract} A wide range of astrophysical sources exhibit extreme and rapidly varying electromagnetic emission indicative of efficient non-thermal particle acceleration. Understanding these sources often involves comparing data with a broad range of theoretical scenarios. To this end, it is beneficial to have tools that enable not only fast and efficient parametric investigation of the predictions of a specific scenario but also the flexibility to explore different theoretical ideas. In this paper, we introduce \tleco, a versatile and lightweight toolkit for developing numerical models of relativistic outflows, including their particle acceleration mechanisms and resultant electromagnetic signature. Built on the Rust programming language and wrapped into a Python library, \tleco\ offers efficient algorithms for evolving relativistic particle distributions and for solving the resulting emissions in a customizable fashion. \tleco\ uses a fully implicit discretization algorithm to solve the Fokker-Planck (FP) equation with user-defined diffusion, advection, cooling, injection, and escape, and offers prescriptions for radiative emission and cooling. These include, but are not limited to, synchrotron, inverse-Compton, and self-synchrotron absorption. \tleco\ is designed to be user-friendly and adaptable to model particle acceleration and the resulting electromagnetic spectrum and temporal variability in a wide variety of astrophysical scenarios, including, but not limited to, gamma-ray bursts, pulsar wind nebulae, and jets from active galactic nuclei. In this work, we outline the core algorithms and proceed to evaluate and demonstrate their effectiveness. The code is open-source and available in the GitHub repository: \href{https://github.com/zkdavis/Tleco} {https://github.com/zkdavis/Tleco}
\end{abstract}





\section{Introduction}
Traditionally, our collective understanding of the cosmos has been primarily derived from observations of the electromagnetic spectrum. As a result, our knowledge of cosmological sources has been constrained to what theories can explain the observed emissions. In many astrophysical sources, such as gamma-ray bursts (GRBs), active galactic nuclei (AGNs), pulsar wind nebulae (PWNe), and many others, the electromagnetic radiation from these sources is understood to result from a distribution of relativistic particles interacting with magnetic field and/or photon fields \citep{Rybicki:1979,Ghisellini_2013}. Thus, to decipher the environments responsible for the observed emission, it is crucial to understand the physical processes affecting the particle distribution and the resultant radiation.

It is common to address this issue by using a kinetic equation, a type of continuity equation that offers a statistical interpretation of particle distributions \citep{Dermer:2009}. This approach allows for the incorporation of well-understood physical mechanisms to observe their effects on the particle distribution and the resulting electromagnetic radiation \citep{Dermer:2009}. An example includes combining our understanding of shocks' ability to accelerate particles to non-thermal energies with our knowledge of synchrotron radiation. Thus, when combined, we demonstrate that the extended spectrum of the Crab Nebula may result from particles being accelerated to ultra-relativistic energies at the termination shock of the pulsar wind. Energetic particles are then cooled via synchrotron radiation \citep{2008volpi}. Creating and testing these models often involves assembling sets of radiation, diffusion, and advection mechanisms, integrating them into a kinetic equation which includes relevant injection or escape terms, and then solving the equation, which may lack an analytical solution.

Here, we provide an open-source, customizable, efficient, and user-friendly solution: \tleco. At the core of this software solution is the evolution of the kinetic equation and self-consistently calculates the resultant emission. Specifically focused on modeling relativistic plasmas in astrophysical settings, this Python package wraps robust algorithms written in Rust to solve the Fokker-Planck (FP) equation, calculate non-thermal emission from synchrotron and inverse Compton scattering, assess the radiative cooling rate well into the Klein-Nishina (KN) regime, and offer many other essential tools for modeling relativistic astrophysical objects, as described in this paper or the accompanying documentation \citep{Tleco2023, Tleco_zenodo}. \tleco\ was originally developed in FORTRAN 95 (previously known a \texttt{Paramo}) and has been used in several publications \citep{2023Combi,2022Davis,2021Rueda-Becerril}. However, for public release, it was decided to create a more user-friendly version packaged in Python that wraps the underlying Rust code. This approach ensures that the end-user needs only to be familiar with Python, enabling easy support across all major operating systems and platforms. Having the underlying code written in Rust, a highly efficient modern language, ensures our capacity for development and improvement with modern tools for the foreseeable future.

\tleco\ is designed not to simulate a specific scenario, such as modeling a GRB afterglow \citep[e.g.,][]{afterglowpy} or using a more comprehensive code for simulating high-energy astrophysical sources \citep[e.g.,][]{lehamoc}, but rather to provide a library that end-users can utilize to build simulations as needed. This flexibility allows end-users to modify or include physical aspects not originally covered by the code base or to remove those deemed unnecessary. Within \tleco, there are several examples already coded that can easily demonstrate how users may construct their own setups using \tleco.

This paper aims to illustrate the core capabilities of the \tleco\ code base and to explain and validate its underlying components. To achieve this, we will divide the paper into four main sections. Section \ref{sec:FP-eq} will detail the FP equation used in \tleco, including its numerical scheme, and conclude with comparisons and convergence testing. In section \ref{sec:radiation}, we describe the algorithms for computing electromagnetic radiation from a distribution and the associated cooling terms, ending this section with comparisons to previous works. In section \ref{sec:examples}, we showcase some of \tleco's capabilities by recreating steady-state blazar scenario from \citep{Katarzynski:2006gh}. Finally, in section \ref{sec:conclusion}, we discuss future applications and development before concluding.

\section{The Fokker-Planck Equation}
\label{sec:FP-eq}
\tleco's primary objective is to simulate the evolution of relativistic particle distributions across various astrophysical scenarios while generating their corresponding radiative signatures. In numerous astrophysical contexts, it is customary to assume an isotropic distribution without spatial dependence, both initially and throughout its evolution. Under these assumptions, the particle distribution's governing equation, which \tleco\ currently solves, given by\footnote{All quantities represent measurements in the comoving frame of the fluid unless explicitly stated otherwise.},
\begin{eqnarray}
    \frac{\partial n(\gamma, t)}{\partial t} & = & \frac{1}{2} \frac{\partial^2}{\partial\gamma^2} \left[ D(\gamma, t) n(\gamma, t) \right] \nonumber \\
    & + & \frac{\partial}{\partial\gamma} \left[ \dot{\gamma}(\gamma, t) n(\gamma, t) \right] \nonumber \\  & + & Q(\gamma, t) - \frac{n(\gamma, t)}{t_{\rm esc}}. 
    \label{eq:FP}
\end{eqnarray}
The FP equation describes the evolution of the particle distribution $n(\gamma, t)$ as function of time ($t$) and the Lorentz factor ($\gamma$). This distribution can be related to the total number of particles at time $t$ inside the volume $V$ by,
\begin{eqnarray}
    N_{total}(t) = V \int^\infty_0 n(\gamma, t) d\gamma.
  \label{eq:total_particle_number}
\end{eqnarray}
The remaining variables in Eq. \eqref{eq:FP} are the diffusion coefficient ($D$), an energy advection term ($\dot{\gamma}$), an injection source term ($Q$), and the parameter ($t_{\text{esc}}$), which describes the eventual particle escape.

\subsection{Discretization}

For the one-dimensional Fokker-Planck equation, many discretization methods have been suggested (for a discussion, see \citep{Park:1996pe}). When choosing a method, its application should be considered carefully. \tleco\ focuses on evolving relativistic particle distributions and producing resultant electromagnetic signatures, tasks that commonly require evaluating many orders of magnitude in both time and energy. This demands an algorithm that remains accurate without requiring unnecessarily small grid spacing. In \citet{Park:1996pe}, several discretization schemes are considered and compared, leading to a general consensus that the scheme proposed by \citet{Chang:1970co} results in the most stable solutions with minimal, if any, sacrifices in accuracy. The \citet{Chang:1970co} scheme is an implicit method that is first-order convergent in time and second-order convergent in energy. This applies to both linear and logarithmic grid spacing. Additionally, this scheme ensures the necessary positive solution for handling probability distributions, incorporates particle conservation (when injection and escape are absent), and adopts a zero flux boundary condition\footnote{For a more detailed discussion of the boundary condition and its implications, see \citet{Park:1996pe,Park1995}}. Following the \citet{Chang:1970co} approach (with considerations from \citet{Park:1996pe}), we obtain the discretized expression,
\begin{widetext}
\begin{eqnarray}
   \frac{n_{j}^{i+1} - n_{j}^{i}}{\Delta t} & = & \frac{1}{\Delta \gamma_{j}} \left\{ \left[ (1 - \delta_{j + \onehalf}) B_{j + \onehalf}^{i} + \frac{1}{\Delta\gamma_{j + \onehalf}} C_{j + \onehalf}^{i} \right]  n_{j + 1}^{i + 1} \right. \nonumber
  \\
  & - & \left[ \frac{1}{\Delta \gamma_{j + \onehalf}} C_{j + \onehalf}^{i} - \delta_{j + \onehalf} B_{j + \onehalf}^{i} - (1 - \delta_{j - \onehalf}) B_{j - \onehalf}^{i} - \frac{1}{\Delta \gamma_{j - \onehalf}} C_{j - \onehalf}^{i} \right] n_{j}^{i + 1} \label{eq:FP-discret}
  \\
  & + & \left. \left[ \frac{1}{\Delta \gamma_{j - \onehalf}} C_{j - \onehalf}^{i} - \delta_{j - \onehalf} B_{j - \onehalf}^{i} \right] n_{j - 1}^{i + 1} \right \} + Q_{j}^{i} - \frac{n_{j}^{i + 1}}{t_{\rm esc}}, \nonumber
  \label{eq:FP-discret}
\end{eqnarray}
where,
\begin{eqnarray}
     B_{j + \onehalf}^{i} = \frac{1}{2} \left( \frac{D_{j + 1}^{i} - D_{j}^{i}}{\Delta \gamma_{j + \onehalf}} \right) + \frac{1}{2} \left( \dot{\gamma}_{j + 1}^{i} + \dot{\gamma}_{j}^{i} \right), & \quad & B_{j - \onehalf}^{i} = \frac{1}{2} \left( \frac{D_{j}^{i} - D_{j - 1}^{i}}{\Delta \gamma_{j - 1 / 2}} \right) + \frac{1}{2} \left( \dot{\gamma}_{j}^{i} + \dot{\gamma}_{j - 1}^{i} \right),
  \\
  C_{j + \onehalf}^{i} = \frac{1}{4} \left( D_{j + 1}^{i} + D_{j}^{i} \right), & \quad & C_{j - \onehalf}^{i} = \frac{1}{4} \left( D_{j}^{i} + D_{j - 1}^{i} \right),
  \\
  w_{j + \onehalf} = \frac{B_{j + \onehalf}^{i}}{C_{j + \onehalf}^{i}} \Delta \gamma_{j + 1 / 2}, & \quad & w_{j - \onehalf} = \frac{B_{j - \onehalf}^{i}}{C_{j - \onehalf}^{i}} \Delta \gamma_{j - \onehalf},
  \\
  \Delta \gamma_{j + \onehalf} =\gamma_{j + 1} - \gamma_{j}, & \quad & \Delta \gamma_{j - \onehalf} = \gamma_{j} - \gamma_{j - 1}.
  \\
  \delta_{j} = \frac{1}{w_{j}} - \frac{1}{\exp(w_{j}) - 1}, & \quad & \Delta \gamma_{j} = \sqrt{\Delta \gamma_{j + \onehalf} \Delta \gamma_{j - \onehalf}}.
\end{eqnarray}
\end{widetext}
Eq.~\eqref{eq:FP-discret} then corresponds to the tridiagonal system,
\begin{eqnarray}\label{eq:FP_tridiagonal}
    a^i_j & = & \frac{\Delta t}{\Delta \gamma_j} \frac{C^i_{j - \onehalf}}{\Delta \gamma_{j - \onehalf}} \frac{w^i_{j - \onehalf}}{\exp\left(w^i_{j - \onehalf}\right) - 1}, \nonumber
    \\
    b^i_j & = & 1 + \frac{\Delta t}{\Delta \gamma_j}  \frac{C^i_{j-\onehalf}}{\Delta \gamma_{j-\onehalf}} \frac{w^i_{j - \onehalf}}{1 - \exp\left(w^i_{j - \onehalf}\right)} \nonumber \\
    & + &  \frac{\Delta t}{\Delta \gamma_j} \frac{C^i_{j+\onehalf}}{\Delta \gamma_{j+\onehalf}} \frac{w^i_{j + \onehalf}}{\exp\left(w^i_{j + \onehalf}\right) - 1} + \frac{\Delta t}{t_{esc}},
    \\
    c^i_j & = & \frac{\Delta t}{\Delta \gamma_{j}}\frac{C^i_{j+\onehalf}}{\Delta \gamma_{j+ \onehalf}}\frac{w^i_{j+\onehalf}}{1 - \exp\left(-w^i_{j+\onehalf}\right)}, \nonumber
    \\
    r^i_j & = & \Delta t\, Q^i_j + n^i_{j}. \nonumber
\end{eqnarray}
Since $w_j$ is susceptible to numerical overflow or underflow, the precautions described in \citet{Park:1996pe} are taken into account. We then solve for $n^{i+1}_{j}$ by using the tridiagonal algorithm described in \citet{Press:2007}.


\subsection{Testing FP Eqution Solver}
\begin{figure}
  \centering
  \begin{tabular}{@{}c@{}}
    \includegraphics[width=\linewidth]{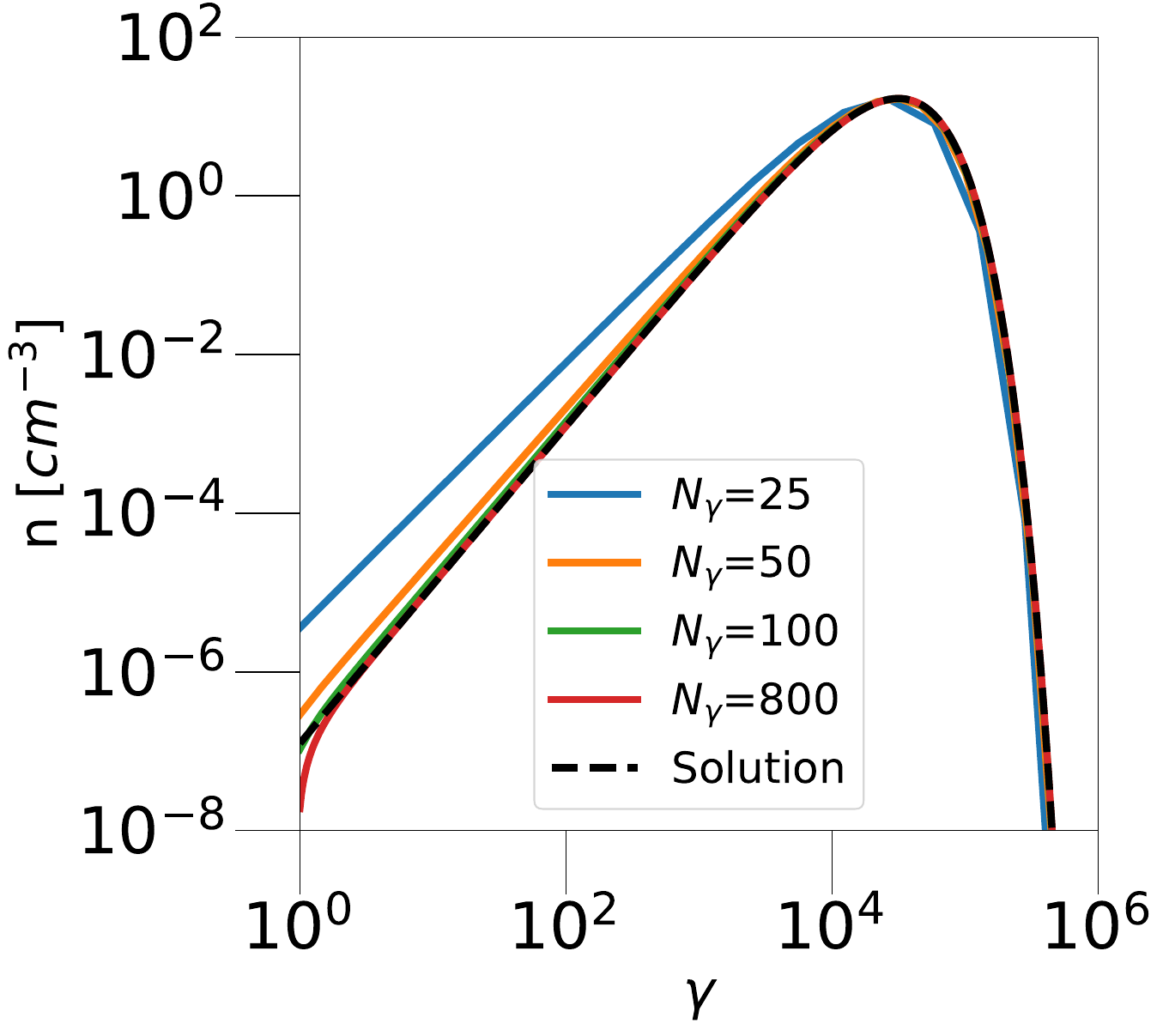} \\[\abovecaptionskip]
  \end{tabular}
  \vspace{-4.25cm}
  \hspace{-1.0cm}
  \begin{tabular}{@{}c@{}}
    \includegraphics[width=\linewidth]{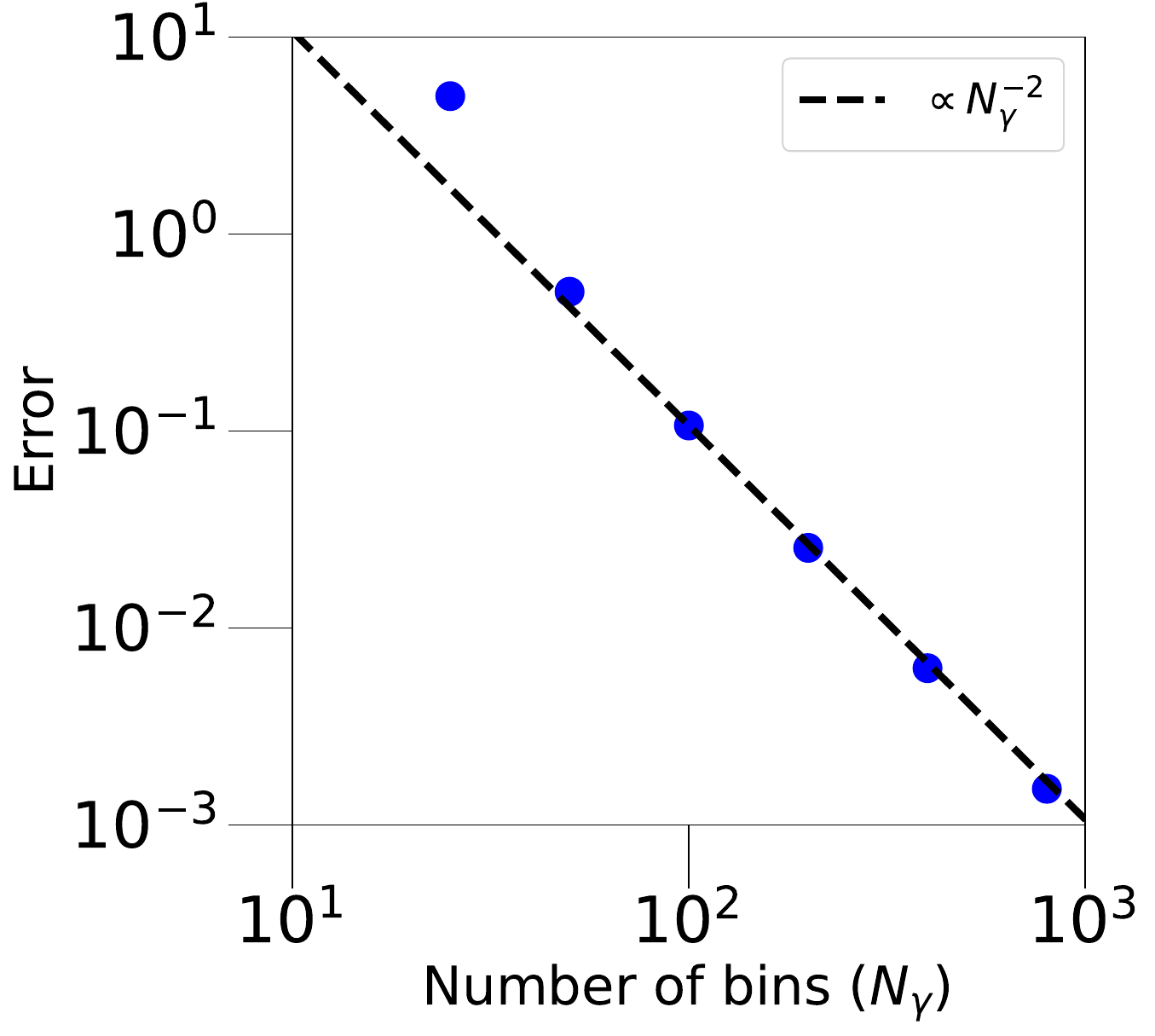} \\[\abovecaptionskip]
  \end{tabular}
  \vspace{4.0cm}
  \caption{Solutions and convergence to a steady state for various values of energy bins ($N_\gamma$). \emph{Top}: solid colored lines show the steady-state distribution for various values of $N_\gamma$. Some values of $N_\gamma$ shown in Tab.~\ref{tab:g_conv_params} have been removed for clarity. Convergence to the analytical solution (black dashed line) is observed as $N_\gamma$ increases. At the low energy boundary, the error $\epsilon$ originates from the singular nature of the FP and the zero flux boundary condition \citep{Park1995}. \emph{Bottom}: The blue points show the error for a given value of N. The black dashed line shows the expected convergence rate from the \citet{Chang:1970co} discretization.}
  \label{fig:steady_state_convergence}
\end{figure}

\begin{figure}
  \centering
  \begin{tabular}{@{}c@{}}
    \includegraphics[width=\linewidth]{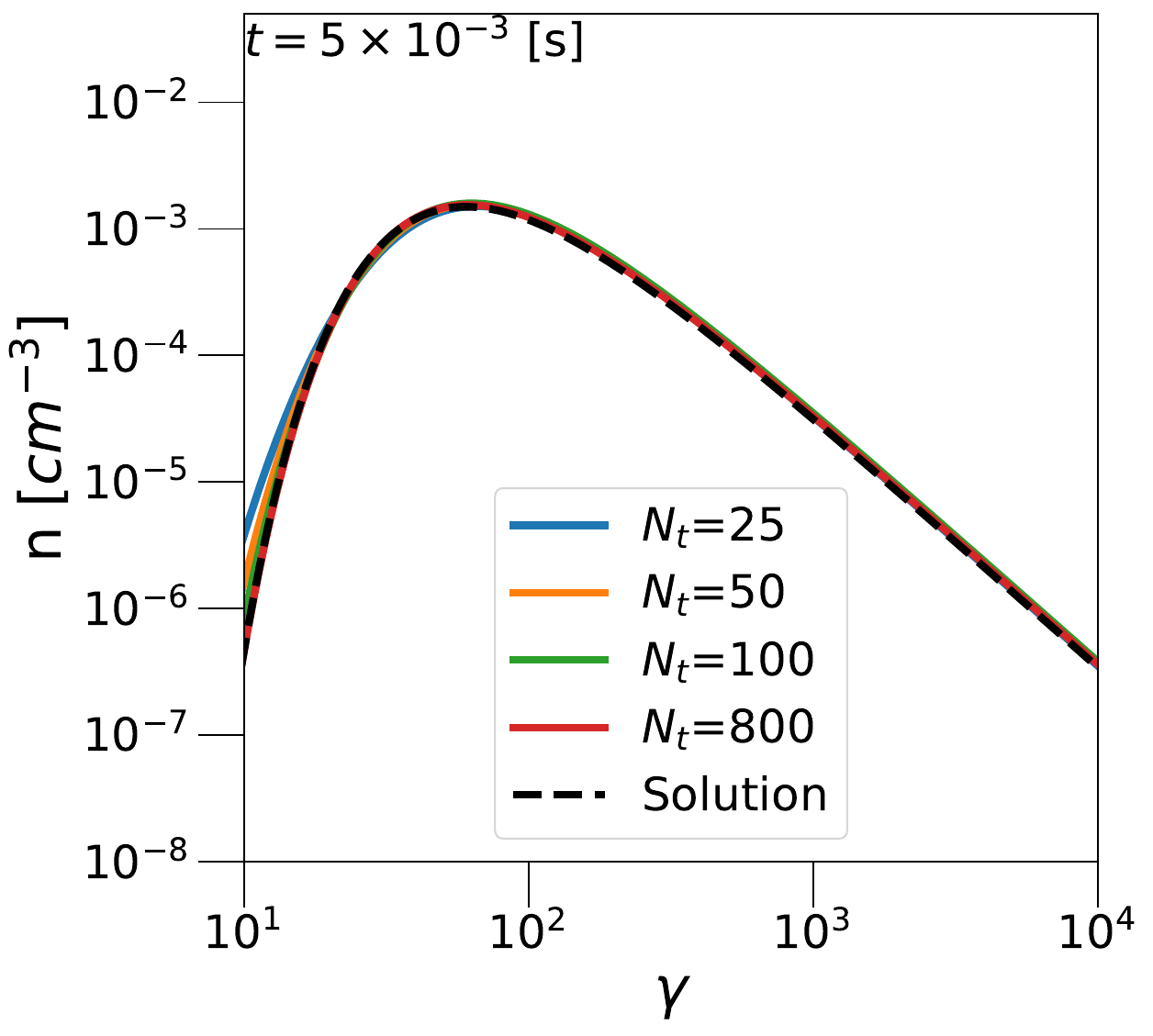} \\[\abovecaptionskip]
  \end{tabular}
  \vspace{-4.25cm}
  \hspace{-1.0cm}
  \begin{tabular}{@{}c@{}}
    \includegraphics[width=\linewidth]{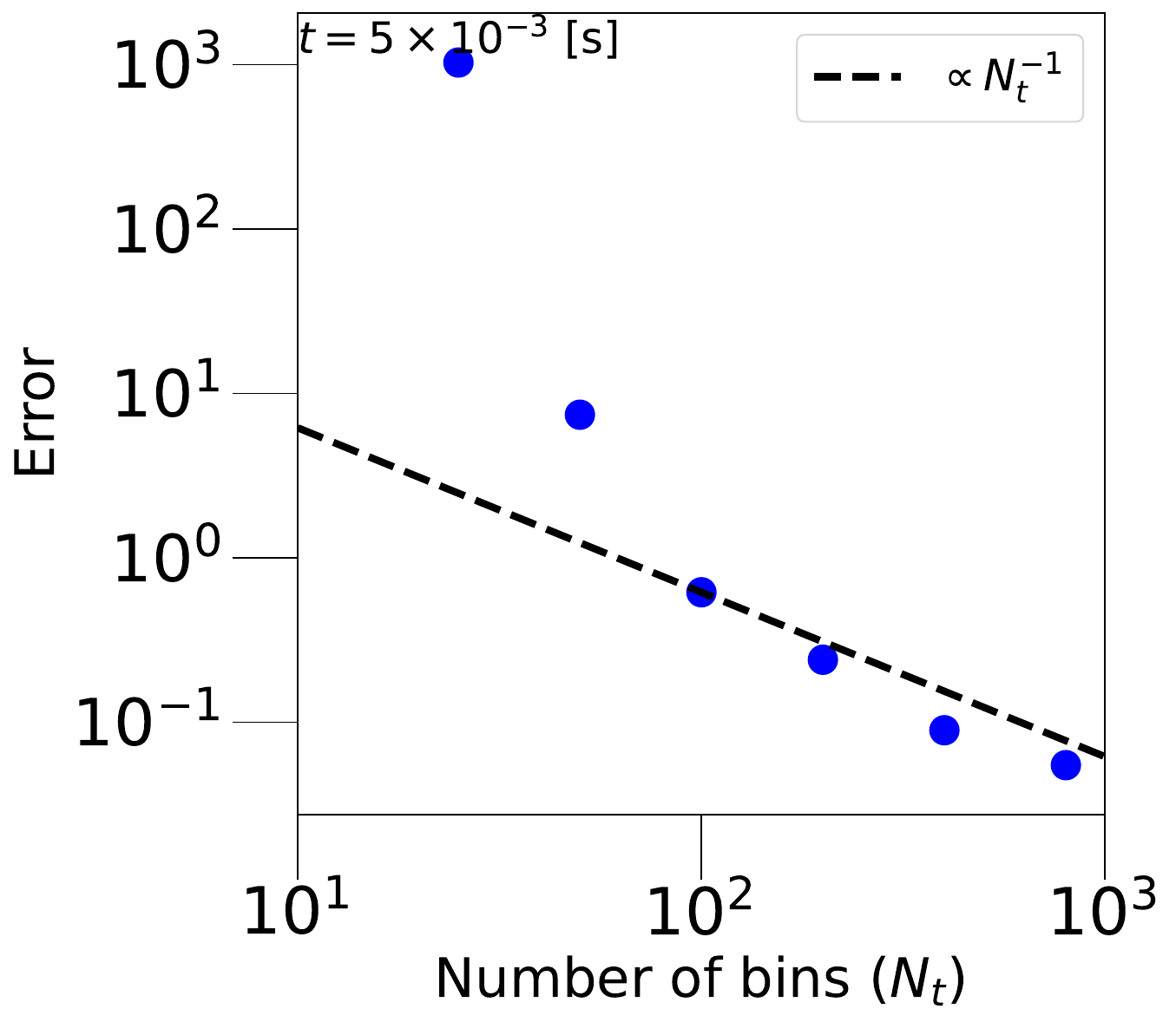} \\[\abovecaptionskip]
  \end{tabular}
  \vspace{4.0cm}
  \caption{Solutions and convergence to the analytical solution at $t=5\times 10^{-3}$s for various values of time bins ($N_t$). \emph{Top}: solid colored lines show thesolution at $t=5\times 10^{-3}$s as a function of the number of bins ($N_t$). Some values of $N_t$ shown in Table \ref{tab:t_conv_params} have been removed for clarity. Convergence to the analytical solution (black dashed line) is shown as $N_t$ increases. \emph{Bottom}: The blue points show the error for a given value of $N_t$. The black dashed line shows the expected convergence rate from the \citet{Chang:1970co} discretization.}
  \label{fig:t_E-03_convergence}
\end{figure}

\begin{figure}
  \centering
  \begin{tabular}{@{}c@{}}
    \includegraphics[width=\linewidth]{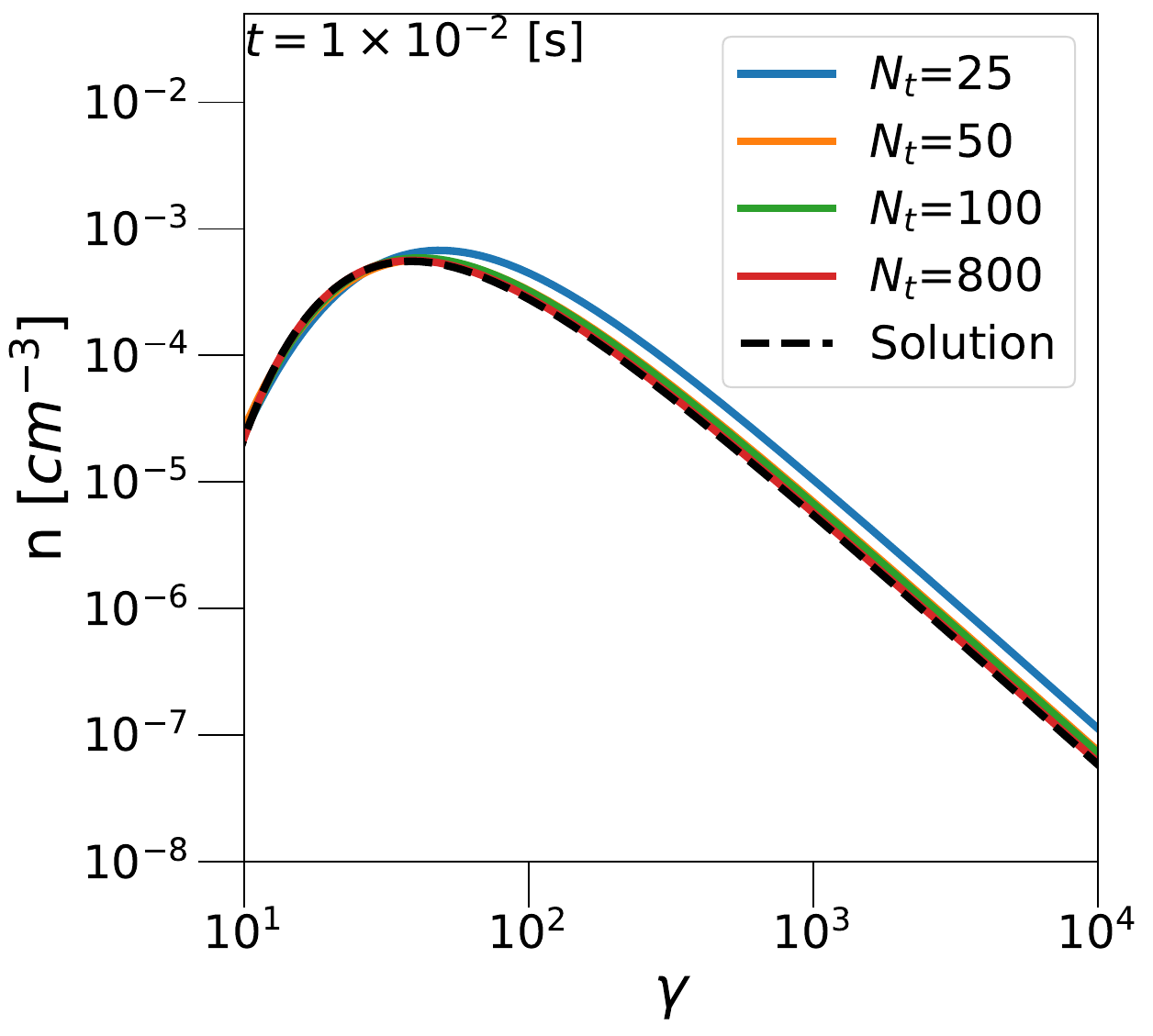} \\[\abovecaptionskip]
  \end{tabular}
  \vspace{-4.25cm}
  \hspace{-1.0cm}
  \begin{tabular}{@{}c@{}}
    \includegraphics[width=\linewidth]{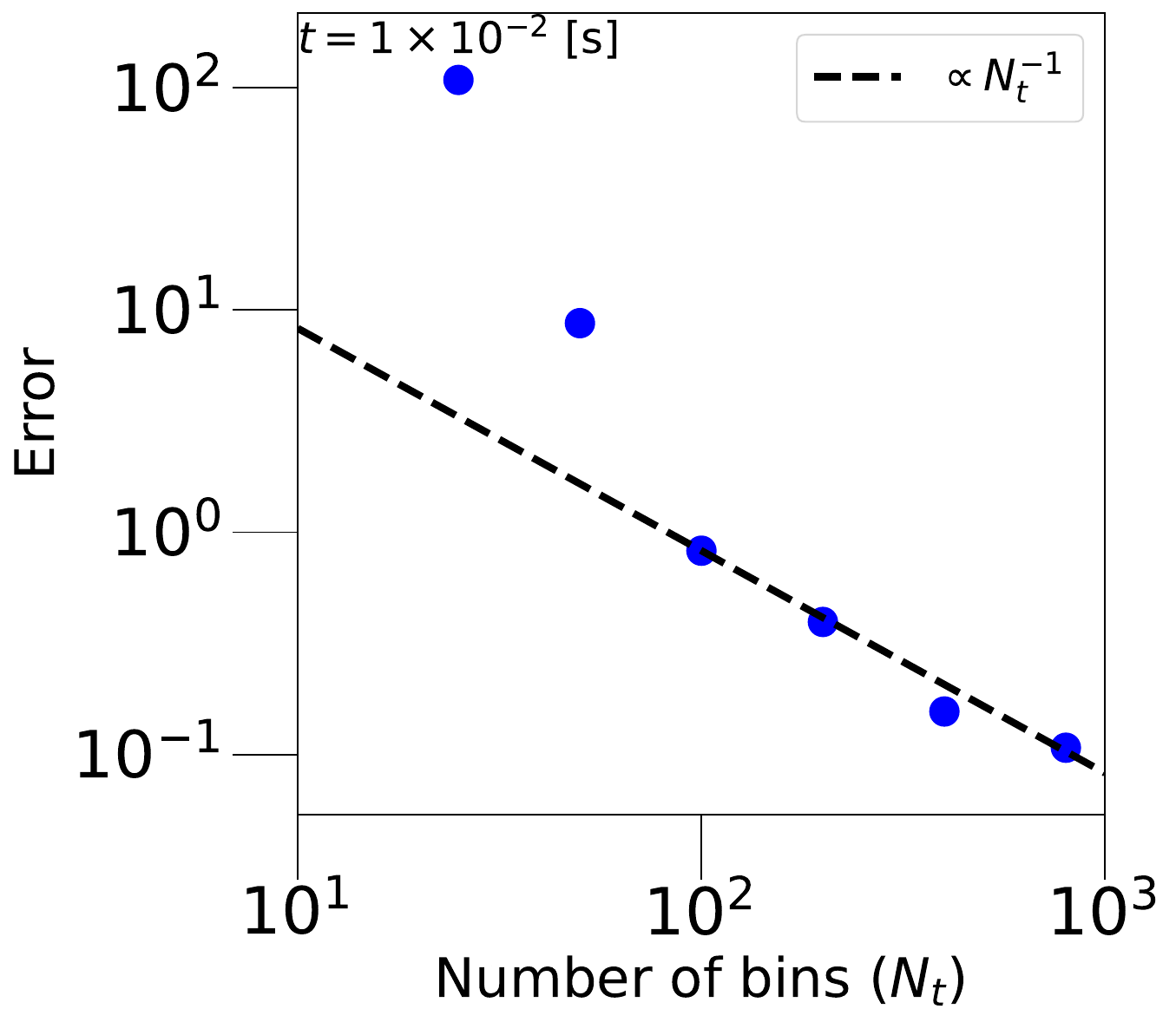} \\[\abovecaptionskip]
  \end{tabular}
  \vspace{4.0cm}
  \caption{Solutions and convergence to the analytical solution at $t=10^{-2}$s as a function of time bins ($N_t$) \emph{Top}: solid colored lines show the solution at $t=10^{-2}$s. Some values of $N_t$ shown in Table \ref{tab:t_conv_params} have been removed for clarity. Convergence to the analytical solution (black dashed line) is shown as $N_t$ increases. \emph{Bottom}: The blue points show the error for a given value of $N_t$. The black dashed line shows the expected convergence rate from the \citet{Chang:1970co} discretization.}
  \label{fig:t_E-02_convergence}
\end{figure}

For code validity, it is essential to ensure that the computed solution matches known analytical solutions, that a variety of FP equations can be solved, and that the algorithm demonstrates expected convergence in both energy and time dimensions. Considering that most applications involve equations that incorporate detailed balance, we accordingly update the FP equation to reflect this,
\begin{eqnarray}
  \frac{\partial n(\gamma, t)}{\partial t}  & = & \frac{\partial}{\partial\gamma} \left[ \frac{1}{2} D(\gamma, t)\frac{\partial}{\partial\gamma} n(\gamma, t)\right] \nonumber
  \\
  & + & \frac{\partial}{\partial\gamma} \left[ \dot{\gamma}_{cool}(\gamma, t) n(\gamma, t) \right],
  \label{eq:FP_detailed_balance}
\end{eqnarray}
by making the substitution,
\begin{equation}
    \label{eq:detailed_balance_sub}
    \dot{\gamma} = \dot{\gamma}_{cool} -\frac{1}{2 \gamma^{2}} \frac{\partial}{\partial\gamma}\left(\gamma^{2} D\right),
\end{equation}
into equation \eqref{eq:FP}\footnote{This should also be done when using \tleco\ if one is using Eq. \eqref{eq:FP_detailed_balance}}. In the above, $\dot{\gamma}_{\text{cool}}$ represents all cooling processes. Focusing first on the algorithm's energy dependence, we examine a setup similar to that described by \citet{Katarzynski:2006gh}, where a diffusion coefficient is chosen to balance radiative cooling, thereby forming a steady state with negligible injection and escape. Following \citet{Katarzynski:2006gh}, we describe the diffusion and cooling terms as $\dot{\gamma}_{\text{cool}} = C_0 \gamma^2$ and $D = 2\gamma_0 C_0 \gamma^2$, respectively. This allows the diffusion to balance the cooling, creating a steady state around $\gamma_{0}$. For this work, we set $C_{0} = 3.48 \times 10^{-11}$ s$^{-1}$ and $\gamma_0 = 10^{4.5}$. $C_0$ corresponds to uniform magnetic field with magnetic energy density value $u_B \approx 0.01$. With these parameters, we can derive a steady-state solution, as described in \citet{Katarzynski:2006gh}, to compare with our computational results.
\begin{equation}
    n_{ss}(\gamma) = \gamma^2 e^{-\frac{2}{\gamma_{0}}(\gamma -1)}
    \label{eq:analytical_solution_1}
\end{equation}
This solution is then used to check convergence with the computational result. For the initial distribution in our tests we used,
\begin{equation}
    \label{eq:injection_powerlaw}
    n(\gamma,t=0) = \left\{
    \begin{array}{lr}
        k \gamma^{-q}, & \text{if } \gamma_{1} \leq \gamma \leq \gamma_{2}\\
        0, & \text{otherwise}
    \end{array}
\right.,
\end{equation}
where $k$ is a constant, $\gamma_{1} = 10^4$, $\gamma_{2} = 10^6$, and the particle distribution index ($q$) is set to 0. The initial distribution is then evolved over a logarithmic time grid with a number of time bins $N_{t} = 800$ from $t=0$ until time $t=200 / \gamma_{0} C_{0} \text{ s}$. This should be about 200 times longer than it takes to form the initial steady state solution. The extra time is to ensure it has formed and that the solution is stable. The number $N_{t}$ is chosen to be large enough to not be a major factor of error (see Figs.\ref{fig:t_E-02_convergence} and \ref{fig:t_E-03_convergence} for time convergence results). Further, the system is solved over a logarithmic energy grid from $\gamma_{min} = 1$ to $\gamma_{max} = 1.5 \times 10^{8}$. The equation is then evolved with a variable number of energy bins $N_{\gamma}$ in order to test the energy convergence rate. This, as well as other parameters in this test, are summarized in table \ref{tab:g_conv_params}.
\begin{table*}[] 
    \centering
    \label{tab:g_conv_params}
    \begin{tabular}{lll}
        Parameter          & Value                      & Description                                \\
        \hline\hline
        $\gamma_{1}$       & $10^{4}$                   & Minimum bound for the initial distribution \\
        $\gamma_{2}$       & $10^{6}$                   & Maximum bound for the initial distribution \\
        $\gamma_{min}$     & 1                          & Smallest energy grid value                 \\
        $\gamma_{max}$     & $1.5 \times 10^{8}$        & Largest energy grid value                  \\
        $N_{\gamma}$ (all) & 25, 50, 100, 200, 400, 800 & All number of energy bins used             \\
        $N_{t}$            & 800                        & Number time bins                           \\
        $C_{0}$            & $3.48 \times 10^{-11}$     & Cooling coefficient                        \\
        $\gamma_{0}$       & $10^{4.5} $                & Mean Lorentz factor                        \\
    \end{tabular}
    \caption{Parameters used to evaluate \tleco's FP energy convergence rate.} 
\end{table*}
In the top frame of Fig. \ref{fig:steady_state_convergence}, it is evident that the code successfully recreates an accurate steady state. In the same figure, for the sake of legibility, we have not included all resolutions tested. The code proves to be extremely accurate even with a small bin size of $N_{\gamma} = 50$, with only minor deviations becoming noticeable in the low energy tail. It is important to note that the trend towards zero in the numerical case is enforced by the zero flux boundary conditions, ensuring that both particles and momentum are conserved. To test the convergence rate, we first define the error,
\begin{equation}
    \label{eq:error}
    \epsilon = \left[ \frac{1}{N}\sum^{N}_{j = 0} \left(\frac{\epsilon^{a}_j - \epsilon^{c}_j}{\epsilon^{a}_j}\right)^2  \right]^{1/2}.
\end{equation}
Here, $\epsilon^{a}_j$ is the analytical solution corresponding to the index $j$, and $\epsilon^c_j$ is the numerical solution. For the convergence plot seen at the bottom of Fig.~\ref{fig:steady_state_convergence}, the error is calculated for the range $10 \leq \gamma \leq 10^{7}$. This is done to avoid obfuscating the convergence rate with errors near the boundary. These errors, due to the singular nature of the FP equation, are discussed in \citet{Park:1996pe}. In the top of Fig. \ref{fig:steady_state_convergence}, this effect can be observed for $N{\gamma} = 800$ at the low energy boundary. In the bottom of Fig. \ref{fig:steady_state_convergence}, we see that the expected convergence is maintained once enough grid points, $N_\gamma \simeq 50$.

To test the time dependence we first add an escape term to our FP,
\begin{eqnarray}
  \frac{\partial n(\gamma, t)}{\partial t} & = & \frac{\partial}{\partial\gamma} \left[ \frac{1}{2} D(\gamma, t)\frac{\partial}{\partial\gamma} n(\gamma, t)\right] \nonumber
  \\
  & + & \frac{\partial}{\partial\gamma} \left[\dot{\gamma}_{cool}(\gamma, t) n(\gamma, t)\right] \nonumber
  \\
  & - & \frac{n(\gamma, t)}{t_{esc}}.
  \label{eq:FP_eq_24}
\end{eqnarray}
We then use terms to match equation 24 in~\citet{Park:1996pe}, making $D=2\gamma^{3}$, $\dot{\gamma}_{cool} = \gamma^{2}$, and $t_{esc} = 1$~s. The time-dependent analytical solution to this equation can be found in~\citet{Park1995}. The analytical solution assumes an initial $\delta$-distribution around $\gamma_{inj}$. Since we can't accurately input a $\delta$-distribution into the computation, we use the analytical solution at time $t=10^{-4}$~s as the initial distribution. The initial distribution is created using $\gamma_{inj} = 100$ and is normalized to 1. For these tests, we again use $\gamma_{min} = 1$ and $\gamma_{max} = 1.5 \times 10^{8}$; the number of energy bins is fixed to a number large enough to not contribute significant error, $N_{\gamma} = 800$, and the bin array is again evenly separated in logarithmic space. The computation runs from time $t_{min} = 10^{-4}$~s to time $t_{max} = 0.1$~s. We then analyze solutions at $\Delta t = t-t_{min} = 5 \times 10^{-3}$~s and $10^{-2}$~s. The time intervals are regular in a logarithmic space grid, and the number of time steps, $N_{t}$, is a variable used to test the time convergence rate. This, along with the other parameters, is summarized in Tab.~\ref{tab:t_conv_params}.

\begin{table*}[]
    \label{tab:t_conv_params}
    \centering
    \begin{tabular}{lll}
        Parameter          & Value                                 & Description                \\
        \hline\hline
        $\gamma_{inj}$     & 100                                   & Injection $\gamma$         \\
        $\gamma_{min}$     & 1                                     & Smallest energy grid value \\
        $\gamma_{max}$     & $1.5 \times 10^{8}$                   & Largest energy grid value  \\
        $N_{\gamma}$       & 800                                   & Number of energy bins used \\
        $N_{t}$ (all)      & 25, 50, 100, 200, 400, 800            & All number time bins       \\
        $t_{min}$          & $10^{-4} $ s                          & Initial time               \\
        $t_{max}$          & $0.1 $ s                              & Max time                   \\
        $\Delta t$         & $5\times 10^{-3}$, $1\times 10^{-2} $ & Evaluation times           \\
        \end{tabular}
    \caption{Parameters used to evaluate \tleco's FP time convergence rate.}
\end{table*}


Figs. \ref{fig:t_E-02_convergence} and \ref{fig:t_E-03_convergence} show the rate of convergence of the numerical solutions to the analytical one as a function of the number of the time bins $N_{t}$. While the accuracy of the numerical solution is sensitively  dependent on the number of time bins, it becomes accurate for $N_{t} \geq 50$ and almost indistinguishable to the theoretical curve for $N_{t} \geq 100$. The error for convergence plots is again calculated using Eq.~\eqref{eq:error} but here we calculate error for all points above the value $\gamma = 10$ for the same reason as mentioned above. The convergence rate trend with $N_{t}$ is less clear for the time dependent solutions but generally holds the expected convergence rate once $N_{t} \simeq 100$.

\section{Radiation and Cooling}
\label{sec:radiation}

Relativistic, charged particles in astrophysical objects may be subject to many radiation mechanisms. These mechanisms produce the electromagnetic signature observed from modeled objects and cool the particle distribution. In this section, we aim to quickly illustrate how electromagnetic radiation is computed from a distribution and how the resultant cooling term can be calculated, and included in the calculations during runtime.

\subsection{Synchrotron Radiation}

When a relativistic, charged particle travels through a magnetic field $B$, it produces synchrotron photons. In \tleco, the synchrotron emissivity is then defined as \citep{Rybicki:1979,Crusius:1988sc}\footnote{From this point forward quantities are assumed to be in the comoving frame unless otherwise stated.},
\begin{equation}\label{eq:emis}
  j_{\nu} = \frac{1}{4 \pi} \int_{1}^{\infty} {\rm d}\gamma P_{\nu}(\gamma) n(\gamma),
\end{equation}
where $P_{\nu}(\gamma)$ is the radiated power spectrum of an electron with Lorentz factor $\gamma$. Currently the emission is only calculated for electrons and positrons.  The power spectrum is assumed pitch angle averaged as described in \citet{Crusius:1986vu},
\begin{equation}
  P_{\nu}(\gamma) = \frac{2 \pi \sqrt{3} {\rm e}^{2} \nu_{\rm g}}{c} F(X_{\rm c}) \text{    },
\end{equation}
where $e$ is the electron's charge, $\nu_{\rm g} = eB/2\pi m_{e} c$ is the gyro frequency, $m_{e}$ is the electron's mass, and $F(X_{\rm c} = 2 \nu / 3 \gamma^{2} \nu_{\rm g})$ is the synchrotron distribution function. The distribution function $F(X_{\rm c})$ used is a piece-wise fit of the pitch angle averaged distribution from \citet{Crusius:1986vu}, developed by \citet{Rueda:2017phd},
\begin{widetext}
\begin{equation}\label{eq:rma-fit}
    F(X_{\rm c}) :=
    \begin{cases}
        1.80842\, X_{\rm c}^{1/3}, & X_{\rm c} < 0.00032
        \\
        \exp\left( A_0 + A_1 \log(X_{\rm c}) + A_2 \log^{2}(X_{\rm c}) + A_3 \log^{3}(X_{\rm c}) A_4 \log^{4}(X_{\rm c}) + A_5 \log^{5}(X_{\rm c}) \right), & 0.00032 \leq X_{\rm c} \leq 0.65
        \\
        \exp\left( B_0 + B_1 \log(X_{\rm c}) + B_2 \log^{2}(X_{\rm c}) + B_3 \log^{3}(X_{\rm c}) + B_4 \log^{4}(X_{\rm c}) + B_5 \log^{5}(X_{\rm c}) \right), & 0.65 < X_{\rm c} \leq 15.58
        \\
        \frac{\pi}{2} \left( 1 - \frac{11}{18 X_{\rm c}} \right) \exp(-X_{\rm c}), &  X_{\rm c} > 15.58
    \end{cases}
\end{equation}
\end{widetext}
where the coefficients $A_i, B_i, i=1, \ldots, 5$ are given in Table \ref{tab:poly-fit-coefs}.

\begin{table}
    \centering
    \begin{tabular}{rrr}
        $i$ & $A_i$ & $B_i$ \\
        \hline\hline
        0 & $-0.78716264$ & $-0.82364552$ \\
        1 & $-0.70509337$ & $-0.83166861$ \\
        2 & $-0.35531869$ & $-0.52563035$ \\
        3 & $-0.06503312$ & $-0.22039315$ \\
        4 & $-0.00609012$ &  $0.01669180$ \\
        5 & $-0.00022765$ & $-0.02865070$ \\
    \end{tabular}
    \caption{Coefficients of function $F(X_{\rm c})$ \citep{Rueda:2017phd}.}
    \label{tab:poly-fit-coefs}
\end{table}

\begin{figure}
    \centering
    \includegraphics[width=\linewidth]{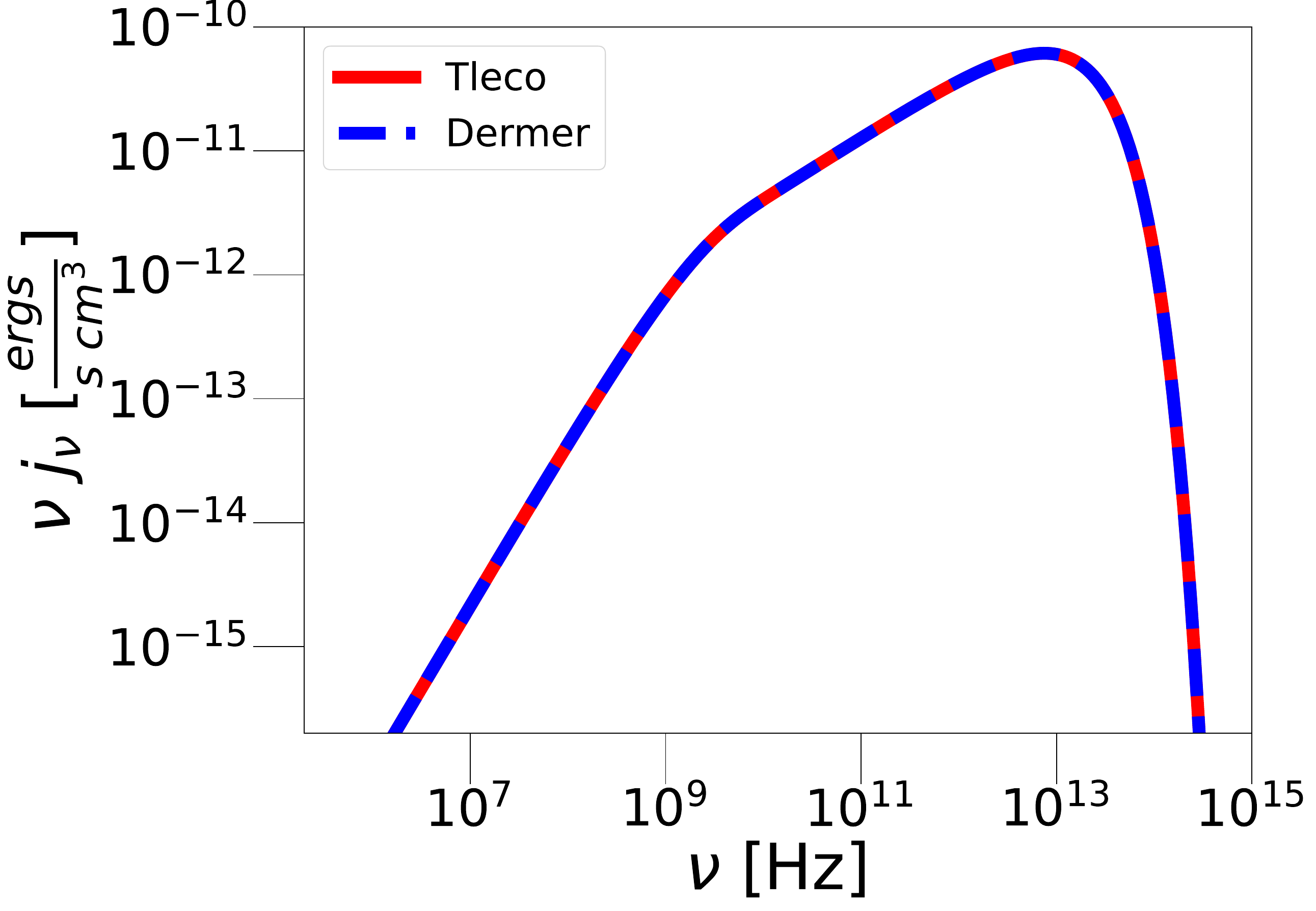}
    \caption{\tleco's synchrotron production comparison. With the solid red line, the \tleco result for a power-law distribution of electrons emitting synchrotron radiation is compared with a solution from \citet{Dermer:2009} (blue dashed lines). For this result, a power index of $2$ is used, with the power-law bounded between $\gamma_1 = 10$ and $\gamma_2 = 10^{3}$. The distribution is normalized to $1$, and the magnetic field is defined so that $u_B = 1 = B^2 / 8\pi$ $\frac{erg}{s  \ cm^3}$. The frequency and Lorentz factor grid are logarithmic, with $N_\gamma = N_\nu = 300$.}
    \label{fig:syn_power_law_comp}
\end{figure}

To calculate the emissivity, we assume the distribution $n(\gamma)$ can be well approximated by a set of power laws the size of the grid $N_\gamma$,
\begin{equation}\label{eq:pwlEED}
  n(\gamma) = \sum^{N_\gamma -1}_{1} n(\gamma_{j}) {\left(\frac{\gamma}{\gamma_{j}} \right)}^{-s} H\left( \gamma; \gamma_{j}, \gamma_{j+1} \right),
\end{equation}
where index $j$ represents a point in the energy grid, $H$ is the Heaviside function, and $n(\gamma_{j})$ is the value of the distribution at index $j$. Substituting the distribution in Eq.~\eqref{eq:pwlEED} into the emission in Eq.~\eqref{eq:emis} gives the final expression for the emissivity,
\begin{equation}\label{eq:emis-pwlEED}
  j_{\nu} = \sum^{N_\gamma -1}_{1} \frac{n(\gamma_{j}) \gamma_{j}^s}{4 \pi} \int_{\gamma_{j}}^{\gamma_{j+1}} {\rm d}\gamma P_{\nu}(\gamma) \gamma^{-s},
\end{equation}
where the final integral is calculated using a standard Romberg integration algorithm \citep{Press:2007}. To illustrate the synchrotron emission, we first compare the computed result with the solution for a power-law particle distribution found in \citet[Eq.~7.51]{Dermer:2009},
\begin{equation}
    \label{eq:dermer_synchrotron}
    j_{\nu} = \frac{\sqrt{3}e^{3}B}{m_{e} c^{2} 4 \pi} \int_{1}^{\infty} \difd\gamma n(\gamma) R(x)
\end{equation}
where $x = \nu / \nu_c$, with $\nu_c \equiv 3 e B \gamma^2 / 4 \pi  m_{e} c$, and
\begin{eqnarray}
    \label{eq:angleaverageddistributionfunction}
    R(x) & = \frac{1}{2} \pi x \left[W_{0,4/3}(x)W_{0,1/3}(x)\right. \nonumber
    \\
    & - \left. W_{1/2,5/6}(x)W_{-1/2,5/6}(x) \right],
\end{eqnarray}
is the angle-averaged synchrotron distribution function \citep{CRUSIUS1986}. $W_{k,u}(x)$ is the Whittaker functions \citep{Abramowitz1972}. The comparison shown in Figure \ref{fig:syn_power_law_comp} utilized a power-law distribution of electrons normalized to $1$, with a power-law index of $2$ between $\gamma_{1} = 10$ and $\gamma_{2} = 10^{3}$. The magnetic field $B$ was chosen so that the magnetic energy density $u_B = B^2 / 8 \pi = 1$. Finally, the energy bin and frequency bin were evenly separated in logarithmic space, with $N_\nu = N_\gamma = 300$.
Figure \ref{fig:syn_power_law_comp} demonstrates that the \tleco\ method for evaluating synchrotron emission is in excellent agreement with the solutions from \citet{Dermer:2009}\footnote{See section 3.2.5 of \citet{Rueda:2017phd}For a full description of the comparison and error calculation of the algorithm}.

\subsection{Absorption}

Once a synchrotron photon is emitted from an accelerated electron in a distribution, it has a non-zero chance of being reabsorbed by another electron in the distribution. Synchrotron self-absorption (SSA), can be summarized for a free electron distribution where the absorbed photon energy is much less than the electron rest mass energy, with the synchrotron absorption coefficient, as explained in sources such as \citep[e.g.,][]{Rybicki:1979,Crusius:1988sc,Ghisellini:1991sv} \citet{Rybicki:1979}, \citet{Crusius:1988sc}, and \citet{Ghisellini:1991sv}.
\begin{align}
    \label{eq:abs}
    \kappa_{\nu} & = -\dfrac{1}{2 m_{e} \nu^2} \int_1^\infty \difd\gamma\, \nonumber
    \\
    & \times \gamma \sqrt{\gamma^2 - 1} P_{\nu}(\gamma) \dfrac{\partial}{\partial \gamma} \left[ \left( \gamma \sqrt{\gamma^2 - 1} \right)^{-1} n(\gamma) \right].
\end{align}
In a similar process to the emission, we assume the electron distribution can be approximated as a series of power-laws, as described by Eq. \eqref{eq:pwlEED}. By incorporating this assumption, we find the absorption coefficient for frequency $\nu$,
\begin{align}
    \label{eq:abs-pwlEED}
    \kappa_{\nu} & = \sum^{N_\gamma -1}_{j=1} \dfrac{n(\gamma_{j}) \gamma_{j}^s}{2 m_{e} \nu^{2}} \int_{\gamma_{j}}^{\gamma_{j+1}} \difd\gamma \nonumber
    \\
    & \times P_{\nu}(\gamma) \gamma^{-(s + 1)} \left[s + 1 + \gamma^2 / (\gamma^2 - 1) \right].
\end{align}
The final integral is calculated using a standard Romberg integration algorithm \citep[see][]{Press:2007}.

\subsection{Compton Emission}

In this section, we set out to explain the implementation of the Compton emission code inside \tleco. A distribution of relativistic electrons incident upon a distribution of photons has the potential to upscatter photons within the distribution. The resultant emissivity in the comoving frame can be given by \citep[see][Eq.~6.60]{Dermer:2009},
\begin{align}
   \label{eq:compton_emission_fund}
   j_{IC}(\nu) & = \frac{h \nu c}{2} \int_0^{\infty} \difd\nu \frac{u(\nu)}{\nu} \nonumber \\
   & \times \int_1^{\infty} \difd\gamma n_e(\gamma) \int_{-1}^1 \difd\mu (1-\mu) \frac{\difd \sigma}{\difd \nu}.
\end{align}
Similarly to the synchrotron approach, to improve computational efficiency, we assume the total spectrum can be created by the sum of emissions from small, power-law approximated segments of the particle distribution, as described by Eq. \eqref{eq:pwlEED}. Additionally, we consider two cases: i) the scattered photons are isotropic and mono-energetic, and ii) the scattered photons come from an isotropic power-law distribution. In the mono-energetic case, we assume a system that experiences a radiation density $u_{ext}(\nu) = u_{0}\delta(\nu-\nu_{ext})$, where $\nu_{ext}$ is the incoming photon frequency. Following \citet[Sec. 2.2]{Mimica2004}, the resultant emissivity in the comoving frame for a power-law segment of the distribution reads,
\begin{equation}\label{eq:monoICemis}
    \begin{aligned}
        &{j_{IC}}^j(\nu)=c \sigma_{\mathrm{T}} u_{0} \nu^{-1}_{\mathrm{ext}} n\left(\gamma_{j }\right) \gamma_{j }^s w^{(1-s) / 2}\\
        &\times
        \begin{cases}
            \begin{aligned}
                &\mathcal{P}^{\mathrm{M} 04}\left(\frac{w}{\hat{\gamma}_{j + 1}^2}, \frac{w}{\gamma_{j}^2}, \frac{s - 1}{2}\right) &
                \\
                & - \mathcal{P}^{\mathrm{M} 04}\left(\frac{w}{\hat{\gamma}_{j+1}^2}, \frac{w}{\gamma_{j }^2}, \frac{s+1}{2}\right), &\frac{1}{4} \leq w \leq \gamma_{j}^2
            \end{aligned}
            \\
            \begin{aligned}
                &\mathcal{P}^{\mathrm{M} 04}\left(\frac{w}{\hat{\gamma}_{j+1}^2}, 1, \frac{s-1}{2}\right) &
                \\
                & - \mathcal{P}^{\mathrm{M} 04}\left(\frac{w}{\hat{\gamma}_{j+1}^2}, 1, \frac{s+1}{2}\right), & \gamma_{j}^{2} < w \leq \hat{\gamma}_{j+1}^2
            \end{aligned}
            \\
            0, \hspace{4.5cm} \text{otherwise}
        \end{cases}
    \end{aligned}
\end{equation}
where $w=\nu / \nu_{ext}$, $\hat{\gamma}_{j} \equiv \min \left(\gamma_{j},\frac{m_{e} c^2}{4 h \nu_{ext}}\right)$ is the effective upper cut-off that limits emission to the Thompson regime\footnote{Upcoming improvements include adding KN regime cooling to be consistent with the KN cooling algorithm discussed in Sec. \ref{sec:rad_cooling}}, and the function $\mathcal{P}^{\mathrm{M} 04}$ is a power-law integral as described in \citet{Mimica2004},
\begin{equation}
    \label{eq:pwlinteg}
    \mathcal{P}^{\mathrm{M} 04}(a,\ b,\ p) \equiv \int^{b}_{a} \difd x x^{p} = \begin{cases}
    \frac{b^{p+1} - a^{p + 1}}{p + 1} & \text{if} \quad p\neq -1 \\
    \log(\frac{b}{a}) & \text{if} \quad p = -1
    \end{cases}
\end{equation}
The total emission is then the sum over the energy bins,
\begin{equation}
    \label{eq:total_mono_ic_emission}
    j_{IC}(\nu) = \sum^{N_{\gamma}-1}_1 {j_{IC}\,}^j(\nu).
\end{equation}
To show the algorithms effectiveness, we compare our result with \citet[Eq.~6.72]{Dermer:2009},
\begin{figure}
    \centering
    \includegraphics[width=\linewidth]{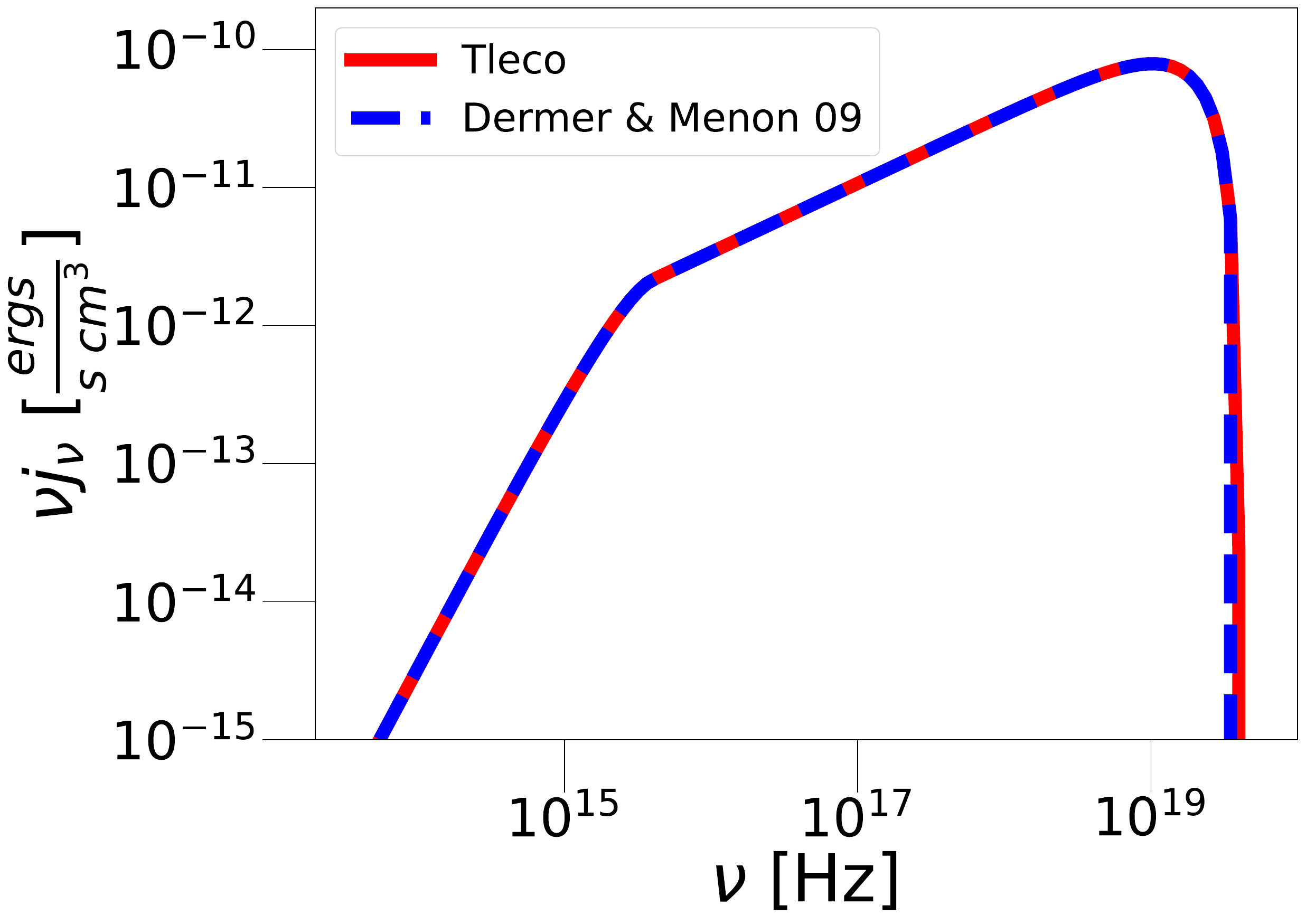}
    \caption{Solid red line compares the \tleco\ result for a power-law distribution of electrons producing Compton emission off a mono-energetic photon field with a solution from \citet{Dermer:2009} (blue dashed lines). A power index of 2 is used, with the power-law bounded between $\gamma_1 = 10$ and $\gamma_2 = 10^{3}$. The distribution is normalized to $1$. The external field is defined so that $u_0 = 1$ and $\nu_{ext} = 10^{13} \ [Hz]$. The frequency and Lorentz factor grid are logarithmic, with $N_\gamma = N_\nu = 300$.}
    \label{fig:IC_mono_comparison}
\end{figure}
For this comparison, the background photon energy density is set to $u_0 = 1\;[\text{ergs}\;\text{cm}^{-3}]$, and the background photon frequency is $\nu_{ext} = 10^{13}\;[\text{Hz}]$. The particle distribution is power-law normalized to a particle density $n_0 = 1\;[\text{cm}^{-3}]$ with an index $p=2$ between $\gamma_1 = 10$ and $\gamma_2 = 10^{3}$. The frequency array ranges from $\nu_{min} = 10^{8}$ to $\nu_{max}=10^{25}$, where both the frequency and Lorentz factor array comprise $300$ bins. Finally, Eq.~(6.72) from \citet{Dermer:2009} is solved using numerical integration. In Figure \ref{fig:IC_mono_comparison}, we observe that \tleco\ replicates the result from \citet{Dermer:2009} with only the slightest deviation at the high-frequency edge.

If the incoming radiation field is not mono-energetic, in the same manner as the prescription for a particle distribution, the incoming intensity in the comoving frame from the external source is approximated as a sum of power-law segments \citep{Mimica2004},
\begin{equation}\label{eq:pwlEED}
  I(\nu) = \sum^{N_\nu -1}_{k=1} I(\nu_{k}) {\left(\frac{\nu}{\nu_{k}} \right)}^{-q} H\left( \nu; \nu_{k}, \nu_{k+1} \right),
\end{equation}
where $N_{\nu}$ is the total number of electromagnetic spectrum frequency bands. Then, for given indices $k$ and $j$, the emissivity is described in \citet{Mimica2004},
\begin{widetext}
\begin{align}\label{eq:pwl_IC_emis_seg}
    j^{j}_{IC\, k}(\nu) & = \sigma_{\mathrm{T}} n(\gamma_j) I(\nu_k)\gamma^s_j(4\nu_k)^q \nu^{-q}\\
    & \times
    \begin{cases}
        G_1^{\mathrm{ISO}}\left(\hat{\gamma}_{j+1}^{-2}, \gamma_{j}^{-2}, w_{k+1}, w_{k}, \frac{s-1}{2}, \frac{2q-s-1}{2}\right), & \frac{1}{4} < \frac{\nu}{4\nu_k} < \gamma_{j}^2
        \\
        G_1^{\mathrm{ISO}}\left(\hat{\gamma}_{j+1}^{-2}, \gamma_{j}^{-2}, w_{k+1}, \gamma_{j}^{2}, \frac{s-1}{2}, \frac{2q-s-1}{2}\right) + G_2^{\mathrm{ISO}}\left(\hat{\gamma}_{j+1}^{-2}, 1, \gamma^{2}_{j}, w_{k}, \frac{s+1}{2}, \frac{2q-s-1}{2}\right), &  \frac{\nu}{4\nu_{k+1}}  \leq  \gamma_{j}^{2} \leq \frac{\nu}{4\nu_{k}}
        \\
        G_2^{\mathrm{ISO}}\left(\hat{\gamma}_{j+1}^{-2}, 1, w_{k+1}, w_{k}, \frac{s-1}{2}, \frac{2q-s-1}{2}\right), & \gamma_{j}^{2} < \frac{\nu}{4\nu_{k+1}} \leq \hat{\gamma}_{j+1}^2
        \\
        0, & \text{otherwise}
    \end{cases}
\end{align}
where $w_{k} \equiv \min \left( \frac{\nu}{4\nu_k}, \hat{\gamma}_{j + 1}^2 \right)$, $w_{k+1} \equiv \max \left( \frac{\nu}{4 \nu_{k + 1}},\frac{1}{4} \right)$. $G^{\mathrm{ISO}}_1$ and $G^{\mathrm{ISO}}_2$ are given by,
\begin{equation}
    \begin{array}{cc}
    G_1^{\mathrm{ISO}} := \mathcal{R}^{M04}(a,b,c,d,\alpha,\lambda) - \mathcal{R}^{M04}(a,b,c,d,\alpha +1,\lambda), \\
    G_2^{\mathrm{ISO}} := \mathcal{S}^{M04}(a,b,c,d,\alpha,\lambda) - \mathcal{S}^{M04}(a,b,c,d,\alpha +1,\lambda). 
    \end{array}
\end{equation}
The functions $\mathcal{R}^{M04}$ and $\mathcal{S}^{M04}$ are described in \citet[Sec.~2.2]{Mimica2004},
\begin{align}
    \mathcal{R}^{M04}(a, b, c, d, \alpha, \beta) & \equiv \int_c^d \difd{x} x^\beta \mathcal{P}^{M04}(x a, x b, \alpha) = \mathcal{P}^{M04}(a, b, \alpha) \mathcal{P}^{M04}(c, d, \alpha + \beta + 1)
    \\
    \mathcal{S}^{M04}(a, b, c, d, \alpha, \beta) & \equiv
    \begin{cases}
        \frac{b^{\alpha+1} \mathcal{P}^{M04}(c, d, \beta)-a^{\alpha+1} \mathcal{P}^{M04}(c, d, \alpha+\beta+1)}{\alpha+1}, & \text{if } \alpha \neq-1 \\
        \log\left(\frac{b}{a}\right) \mathcal{P}^{M04}(c, d, \beta)-\mathcal{Q}^{M04}(c, d, \beta), & \text{if } \alpha=-1
    \end{cases}
\end{align}
where,
\begin{equation}
    \mathcal{Q}^{M04}(a, b, p) \equiv \int_a^b \difd{x} x^p \log(x) =
    \begin{cases}
        \frac{b^{p+1} \log(b) - a^{p+1} \log(a) - P(a, b, p)}{p+1}, & p \neq -1 \\
        \frac{1}{2} \log(a b) P(a, b, p), & p = -1
    \end{cases}
\end{equation}
\end{widetext}

Thus, the total emissivity from the power-law source can be computed as the double sum of Eq. \eqref{eq:pwl_IC_emis_seg} over the energy and frequency bins,
\begin{equation}
    \label{eq:total_pwl_ic_emission}
    j_{IC}(\nu) = \sum^{N_{\nu} - 1}_{j=1} \sum^{N_{\gamma} - 1}_{k=1} {j_{IC\,}}^j_k(\nu).
\end{equation}
As an illustration of the effectiveness of the method, we compare our computation method with the solution from equation 6.89 in \citet{Dermer:2009}. This example is run identically to the mono-energetic comparison, except that the energy density is now a blackbody with a temperature corresponding to a peak at 2~eV.

\begin{figure}
    \centering
    \includegraphics[width=\linewidth]{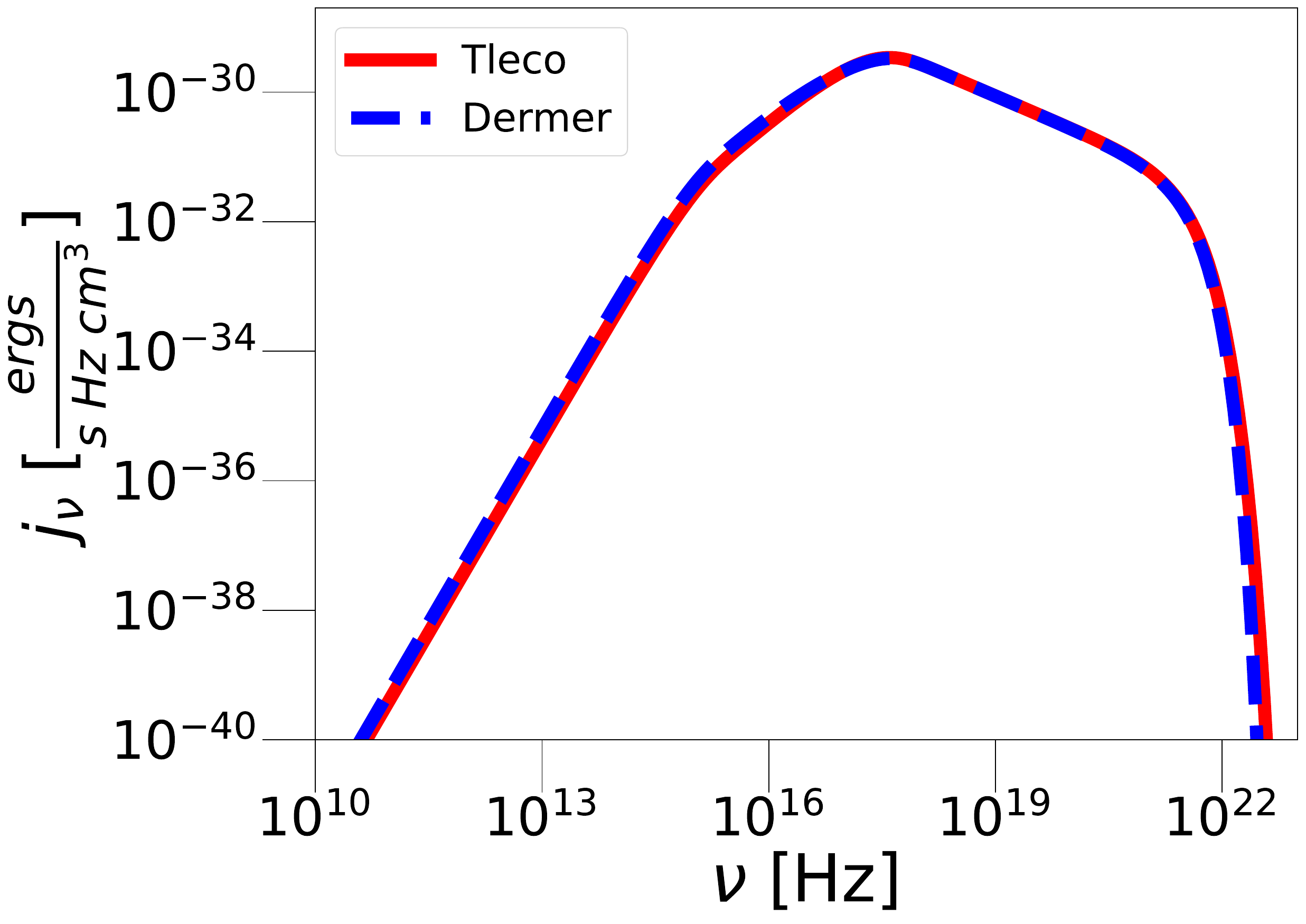}
    \caption{Solid red line compares the \tleco\ result for a power-law distribution of electrons producing Inverse Compton emission off a black-body photon field with a solution from \citet{Dermer:2009} (blue dashed lines). A power index of 2 is used with the power-law bounded between $\gamma_1 = 10$ and $\gamma_2 = 10^{3}$. The distribution is normalized to $1$. The external field is normalized to $u_0 = 1$, and the peak temperature corresponds to photons at 2~eV. The frequency and Lorentz factor grid are logarithmic, with $N_\gamma = N_\nu = 300$.}
    \label{fig:IC_bb_comparison}
\end{figure}
In Fig. \ref{fig:IC_bb_comparison}, the computed solution generally agrees with the result from \citet{Dermer:2009}, with small differences shown below the peak.

\subsubsection{SSC emission}

\tleco\ has no direct implementation of synchrotron self-Compton (SSC) emission. However, a common method of calculating SSC emission involves solving for the synchrotron intensity via the radiation transfer equation. \tleco\ includes several solutions to the radiation transfer equation, and more complex geometries can be implemented, but for this work, we only discuss the slab approximation in section \ref{sec:examples}. The intensity from the synchrotron spectrum can then be used in the same way as the broken power-law assumption made for equation \eqref{eq:total_pwl_ic_emission}. This assumes that any changes to the synchrotron spectrum, and thus photon energy density that results in SSC emission, fluctuate on timescales smaller than the time steps used and fill the blob volume. An example of this is carried out in Sec.~\ref{sec:examples}.

\subsection{Radiative Cooling}
\label{sec:rad_cooling}
In order to correctly evolve the emission from a relativistic, charged particle distribution, one needs to accurately account for the cooling effect the emission has on the distribution. In this section, we will focus on the numerical treatment for synchrotron and Compton cooling. For processes such as synchrotron, the result can be found in \citet{Rybicki:1979},
\begin{equation}
    \label{eq:sync_cooling}
    \dot{\gamma}_{syn} (\gamma) = \frac{4}{3} \sigma_T c \frac{u_B}{m_e c^2} \gamma^2 ,
\end{equation}
Compton cooling is often found to be dynamic and non-linear but can be expressed as \citep{Rybicki:1979},
\begin{equation}
    \label{eq:compton_cooling}
    \dot{\gamma}_{C} (\gamma,t) = \frac{4}{3} \sigma_T c \frac{u_{rad}(\gamma,t)}{m_e c^2} \gamma^2 .
\end{equation}
The radiation energy density, $u_{rad}$, is composed of two sources: a pre-described external energy density, $u_{ext}$, due to a photon field in the surrounding environment responsible for inverse Compton, and the energy density of the synchrotron photons emitted by the accelerated particles, $u_{syn}$ \citep{Rybicki:1979}.
\begin{equation}
    \label{eq:u_rad}
    u_{rad}(\gamma,t) = u_{ext}(\gamma,t) + u_{syn}(\gamma,t).
\end{equation}
The synchrotron photon density is given by the synchrotron intensity \citep{Rybicki:1979},
\begin{equation}
    \label{eq:u_rads}
    u_{syn}(\gamma, t) = \frac{4\pi}{c} \int^{\nu_{max}(\gamma)}_{\nu_{min}} \difd \nu I_{\nu}(t),
\end{equation}
where $\nu_{max}(\gamma) = \min\left[\nu_{max}, m_e c^2 / h \gamma \right]$ and the integral in eq.~\eqref{eq:u_rads} is approximated using power-law segments as done previously,
\begin{equation}
    \label{eq:u_rad3}
    u_{syn}(\gamma, t) \approx \frac{4\pi}{c} \sum^{N_{\nu} - 1}_{1} I_{\nu,i} \nu_{i}^{\alpha_i} \mathcal{P}^{\mathrm{M}04}\left(\frac{\nu_{i+1}}{\nu_{i}},\alpha_{i}\right).
\end{equation}
Eqs. \eqref{eq:compton_cooling} and \eqref{eq:u_rad3} are effective as long as we are in the Thompson scattering regime. Once in the KN regime, it's necessary to account for the reduced cooling efficiency. This reduced efficiency is due to QED corrections to the scattering cross section when $\gamma h \nu > m_e c^2$. \tleco\ includes an implementation of KN cooling that is tested and described in \citet{Rueda2021kn}.

\section{Example: Steady States via Stochastic Particle Acceleration and Radiative Cooling}
\label{sec:examples}

In this section, we aim to illustrate \tleco's capability to numerically produce particle distributions and the resulting radiation for various astrophysical situations by reproducing, as closely as possible, the scenarios outlined in \citet{Katarzynski:2006gh}. The goal of \citet{Katarzynski:2006gh} is to understand the effects of stochastic particle acceleration on a particle distribution and its eventual radiation. To achieve this, \citet{Katarzynski:2006gh} initially sets up 6 steady-state scenarios using the FP prescription.
\begin{equation}
    \label{FP:katarzynski_2006}
    \begin{aligned}
        \frac{\partial n(\gamma, t)}{\partial t}  = & \frac{\partial}{\partial\gamma} \left[ \frac{1}{2} D(\gamma, t)\frac{\partial}{\partial\gamma} n(\gamma, t) \right.
        \\
        + & \left.\dot{\gamma}_{cool}n(\gamma, t) \right]
        \\
        - & \frac{n(\gamma, t)}{t_{esc}} + Q(\gamma, t),
    \end{aligned}
\end{equation}
where $Q(\gamma,t)$ is the injection profile. For these examples, $Q(\gamma,t) = n(\gamma, t=0)/t_{inj}$ always uses the initial particle distribution injected over a time $t_{inj}$\footnote{Details on the exact value of $t_{inj}$ are not provided in \citet{Katarzynski:2006gh}, and its value had to be estimated from the plots in the paper.}. The setup assumes that particles are accelerated in situ with $D(\gamma, t) = \gamma^2 / t_{acc}$, where $t_{acc}$ is the diffusive acceleration time scale. For cooling, $\dot{\gamma}_{cool} = C_0 \gamma^2$, and $C_0$ is the cooling constant set to a constant value of $C_0 = 3.48 \times 10^{-11}$. The distributions are cooled to an equilibrium energy $\gamma_e = 10^{4} = 1/t_{acc} C_{0}$. All setups are initially injected with a flat power-law distribution.
\begin{equation}\label{eq:injection_powerlaw}
    n(\gamma,t=0) = \left\{
    \begin{array}{lr}
        n_0, & \text{if } \gamma_{1} \leq \gamma \leq \gamma_{2}\\
        0, & \text{otherwise}
    \end{array}
    \right. .
\end{equation}
The first set of simulations focuses solely on the electron distribution. To summarize the list of parameters for each setup, we list them in Table \ref{tab:steady_state_1_params}.
\begin{table}[]
    \centering
    \label{tab:steady_state_1_params}
    \begin{tabular}{llllll}
    Figure       & $n_0$ & $\gamma_1$ & $\gamma_2$        & $t_{inj}$               & $t_{esc}$ \\
    \hline\hline
    Figure \ref{fig:katazynski_2006_fig1} (top) & $1$        & $1$        & $2$               & $\infty$                & $\infty$  \\
    Figure \ref{fig:katazynski_2006_fig1} (middle) & $1$        & $10^{6}$   & $2 \times 10^{6}$ & $\infty$                & $\infty$  \\ 
    Figure \ref{fig:katazynski_2006_fig1} (bottom) & $1$        & $10$       & $10^{6}$          & $\infty$                & $\infty$  \\ 
    Figure \ref{fig:katazynski_2006_fig2} (top) & $1$        & $1$        & $2$               & $t_{acc} \times 10^{4.4}$ & $t_{acc}$   \\ 
    Figure \ref{fig:katazynski_2006_fig2} (middle) & $1$        & $10^{6}$   & $2 \times 10^{6}$ & $t_{acc} \times 10^{4.4}$ & $t_{acc}$   \\ 
    Figure \ref{fig:katazynski_2006_fig2} (bottom) & $1$        & $10$       & $10^{6}$          & $t_{acc} \times 10^{4.4}$ & $t_{acc}$   \\ 
    \end{tabular}
    \caption{Parameters used for Figs. \ref{fig:katazynski_2006_fig1} and \ref{fig:katazynski_2006_fig2}}
\end{table}

\begin{figure}
  \centering
  \begin{tabular}{@{}c@{}}
    \includegraphics[width=\linewidth]{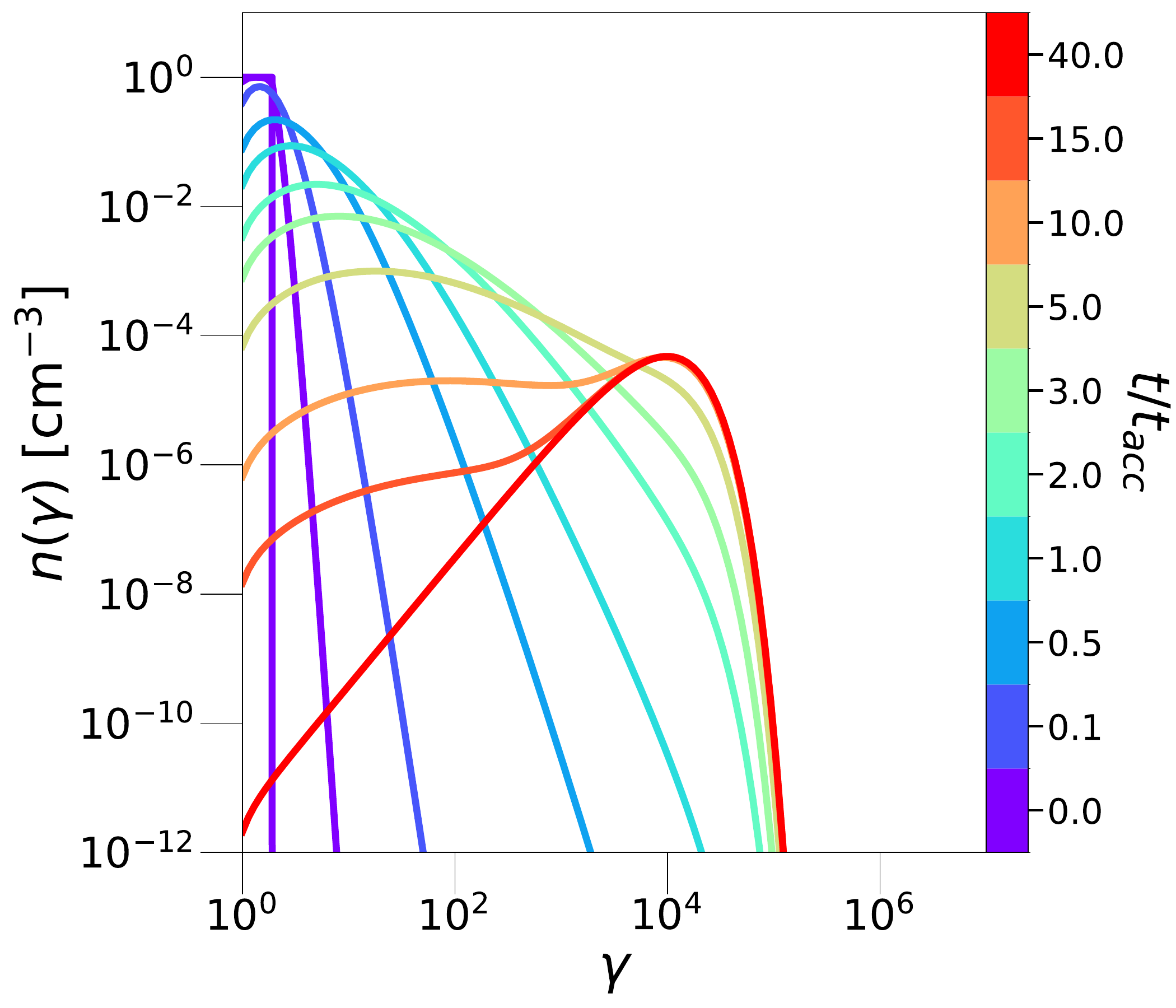} \\[\abovecaptionskip]
  \end{tabular}
  \hspace{-1.0cm}
  \begin{tabular}{@{}c@{}}
    \includegraphics[width=\linewidth]{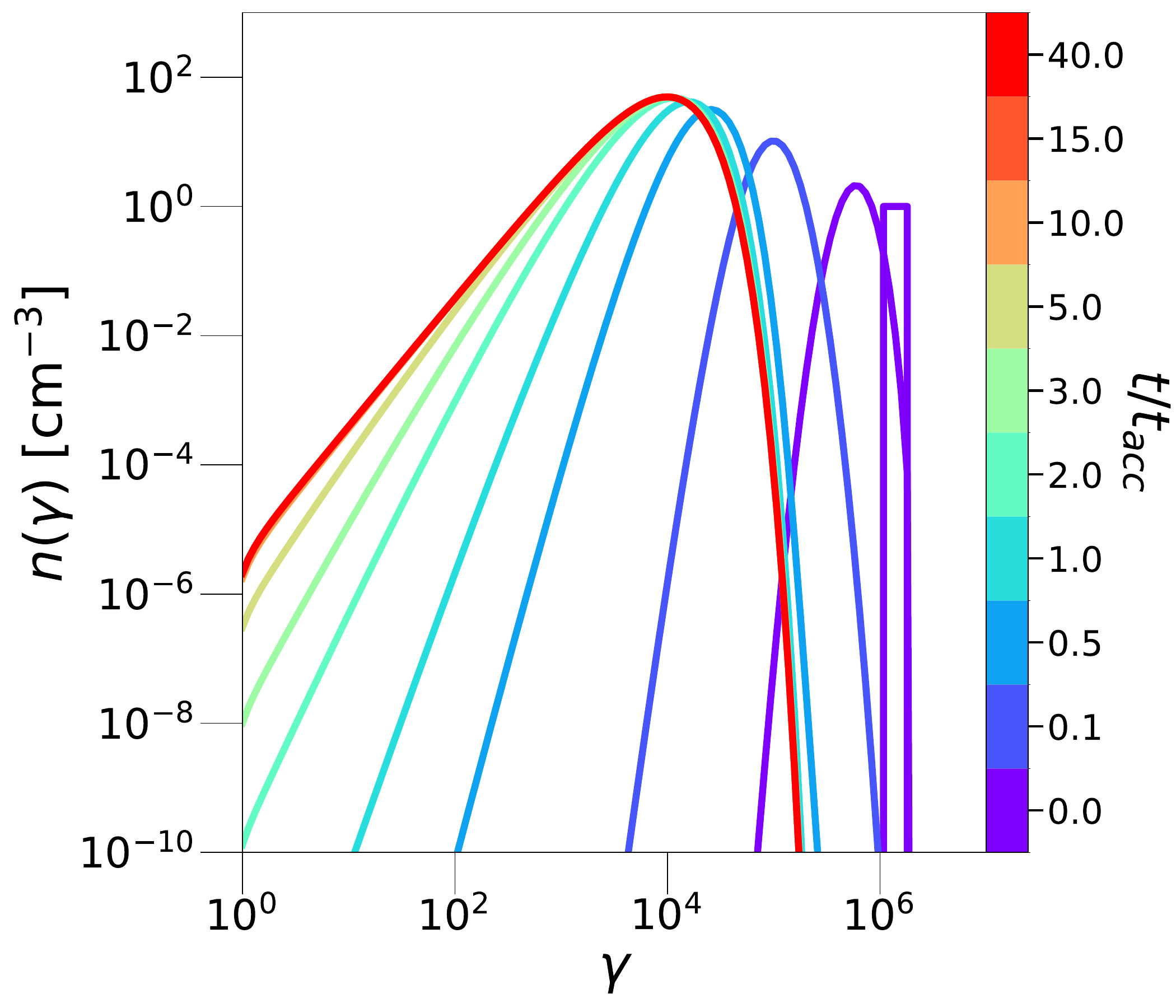} \\[\abovecaptionskip]
  \end{tabular}
  \vspace{-4.25cm}
  \hspace{-1.0cm}
  \begin{tabular}{@{}c@{}}
    \includegraphics[width=\linewidth]{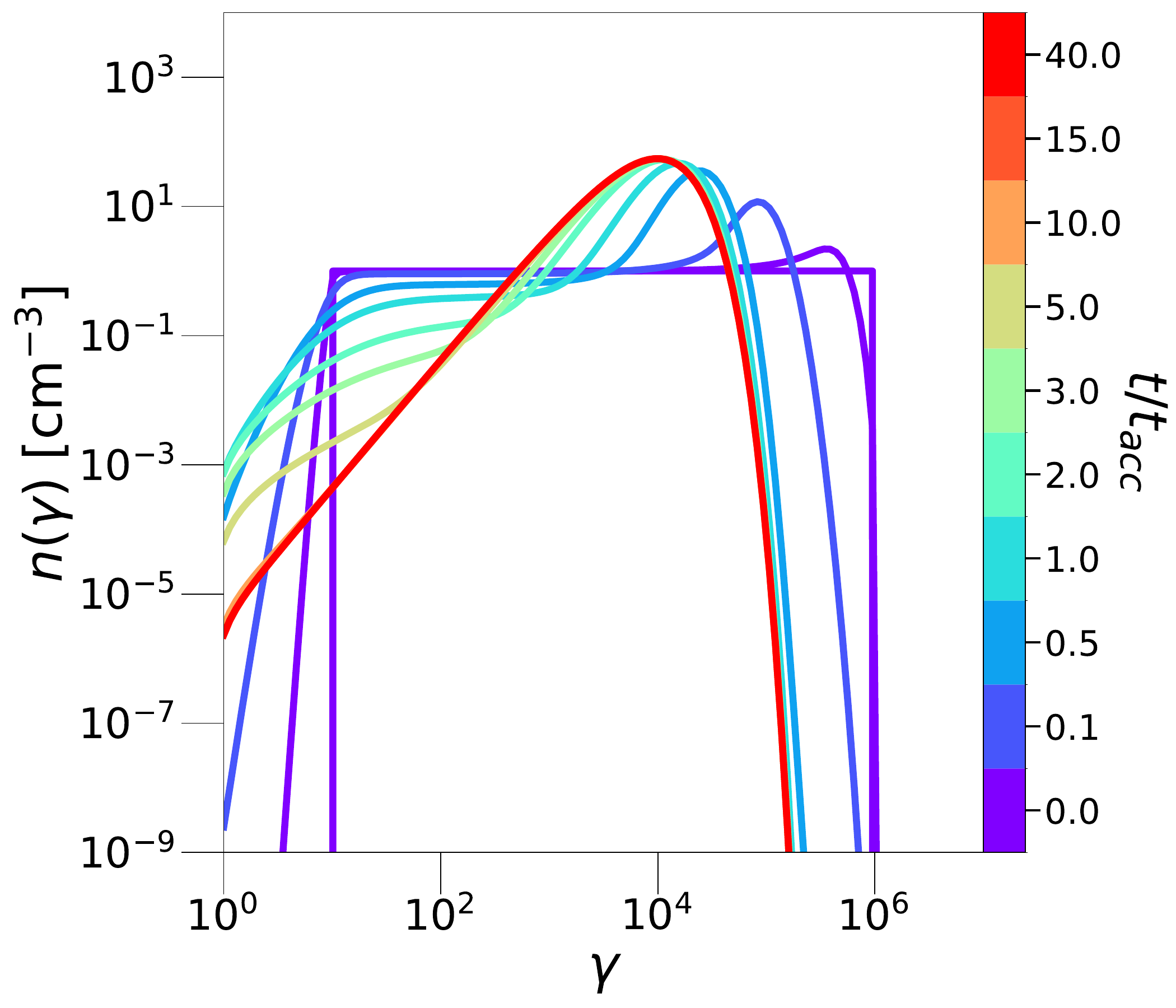} \\[\abovecaptionskip]
  \end{tabular}
  \vspace{4.0cm}
  \caption{For three different initial conditions, top, middle, and bottom (summarized in Table \ref{tab:steady_state_1_params}),convergence to a similar steady state dictated by the balance between stochastic acceleration and radiative cooling is demonstrated. This is consistent with what is shown in \citet{Katarzynski:2006gh}.}
  \label{fig:katazynski_2006_fig1}
\end{figure}

\begin{figure}
  \centering
  \begin{tabular}{@{}c@{}}
    \includegraphics[width=\linewidth]{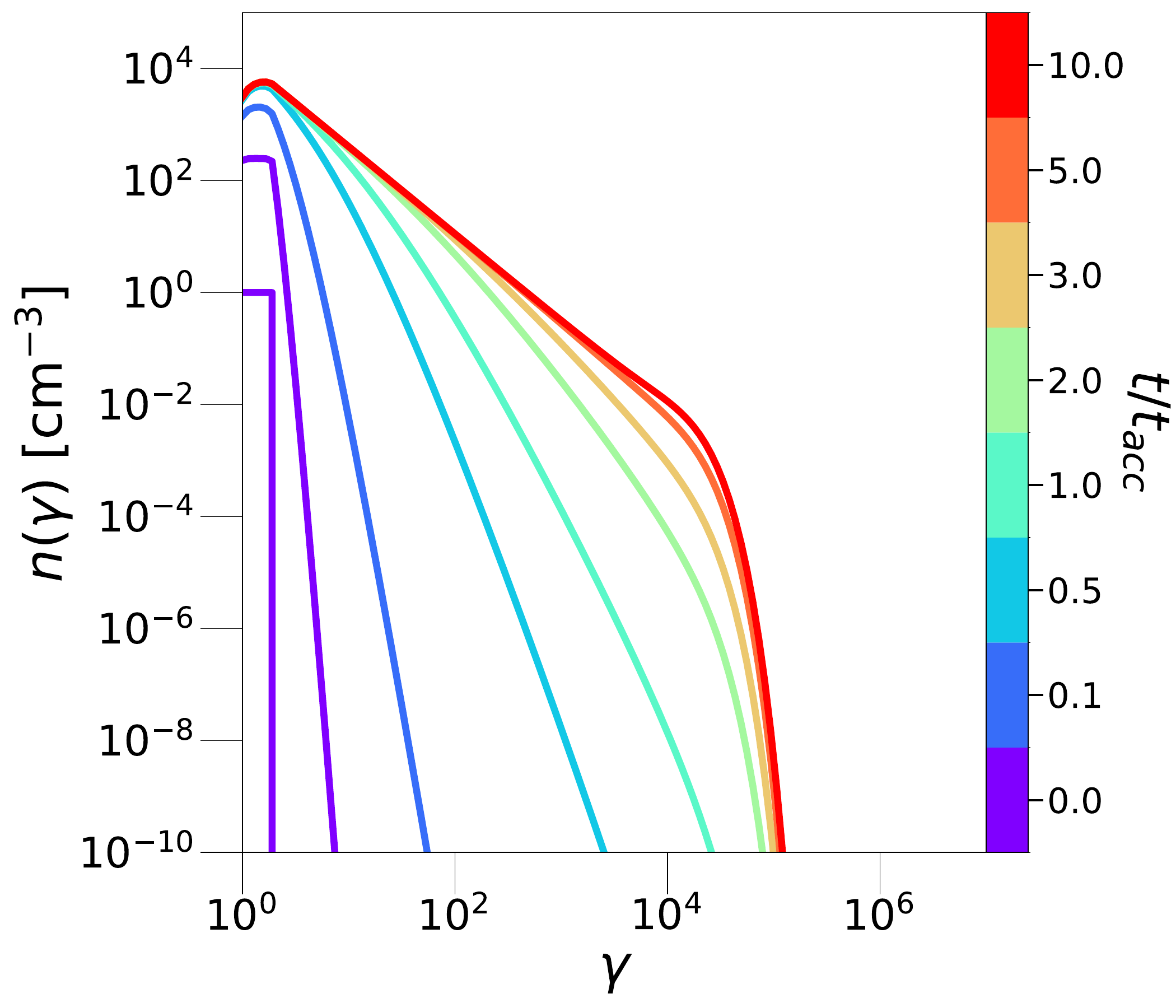} \\[\abovecaptionskip]
  \end{tabular}
  \hspace{-1.0cm}
  \begin{tabular}{@{}c@{}}
    \includegraphics[width=\linewidth]{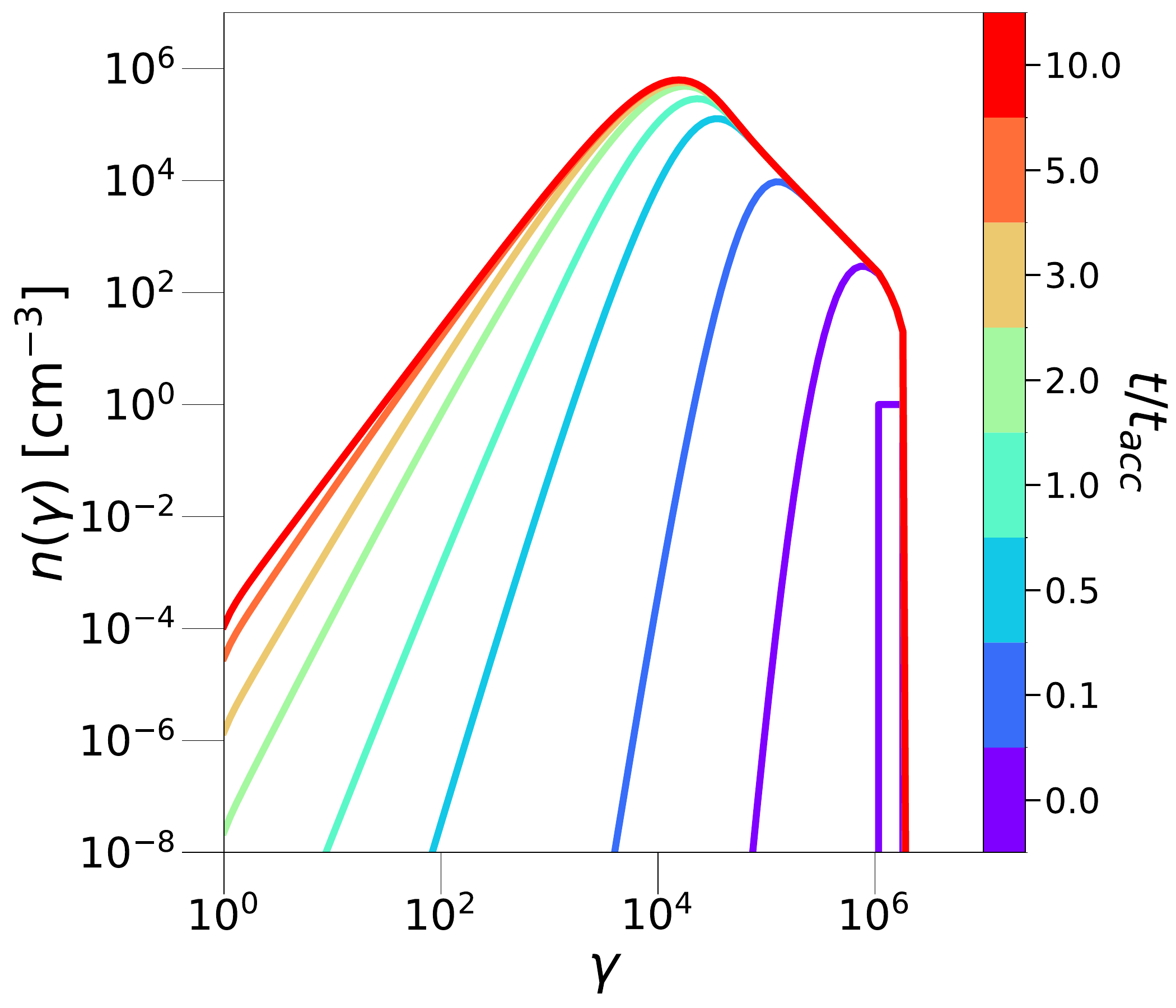} \\[\abovecaptionskip]
  \end{tabular}
  \vspace{-4.25cm}
  \hspace{-1.0cm}
  \begin{tabular}{@{}c@{}}
    \includegraphics[width=\linewidth]{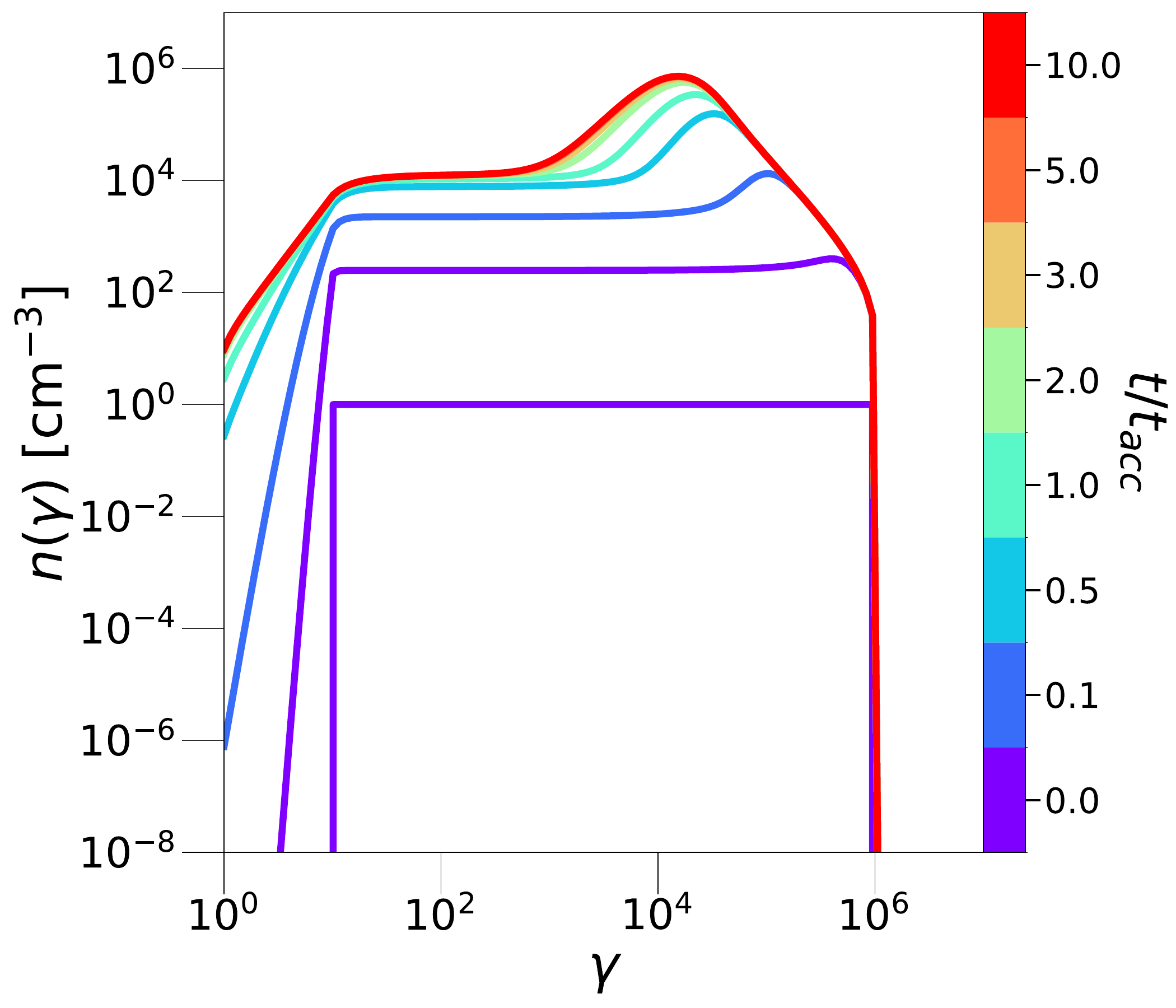} \\[\abovecaptionskip]
  \end{tabular}
  \vspace{4.0cm}
  \caption{For three different initial conditions, top, middle, and bottom (summarized in Table \ref{tab:steady_state_1_params}), it is shown how injection and escape can modify the shape of the steady state. This is consistent with what is shown in \citet{Katarzynski:2006gh}.}
  \label{fig:katazynski_2006_fig2}
\end{figure}

Although a similar steady state was already tested (see Fig.~\ref{fig:katazynski_2006_fig1}) in Sec.~\ref{sec:FP-eq}, the more complex solutions that come from adding injection and escape (Fig.  \ref{fig:katazynski_2006_fig2}) also agree with the works presented in \citet{Katarzynski:2006gh}. There is a slight difference at the low energy boundary that comes from the zero flux boundary condition discussed in Section \ref{sec:FP-eq}.

The following examples from \citet{Katarzynski:2006gh} are modified to simulate the emission from the highly variable blazar Mrk~501. Here, we do not directly compare to emission and wish only to recreate the work as an example. First, the acceleration and escape time scales are defined by the dynamical time scale, i.e., $t_{acc} = t_{esc} = R/c$, where $R$ is the size of the emitting region. To include self-consistent cooling with the radiation,
\begin{equation}
    \label{eq:gcool_fig3_and_fig4}
    \dot{\gamma}_{cool} =\frac{4}{3} \frac{\sigma_{\mathrm{T}} c}{m_{\mathrm{e}} c^2} \gamma^2 u_B + \dot{\gamma}_{KN},
\end{equation}
where $\dot{\gamma}_{KN}$ is the cooling rate from inverse Compton scattering, calculated using the synchrotron photon intensity ($I_s$) and \tleco's IC cooling with KN corrections as outlined in \citet{Rueda2021kn}. To find $I_s$, we use the radiation transfer approximation of a slab, chosen because it recreates results closest to \citet{Katarzynski:2006gh}, though they do not specify their exact radiation transfer method. Finally, to calculate the flux\footnote{In a future release, we plan to add the ability to calculate the energy flux by a cosmologically distant observer from multiple regions while considering causality.} in the observer frame, we apply a Doppler boost with factor $\delta_D$, and use \citet{Kneiske2010} via \citet{meyer2022} to account for absorption due to extragalactic background light of TeV photons\footnote{This is slightly different from the prescription used in \citet{Katarzynski:2006gh} but was chosen due to being a more updated model of the electromagnetic background, while being easier to implement in this example.}. The rest of the parameters are summarized in Table \ref{tab:steady_state_2_params}.

\begin{table}
    \centering
    \begin{tabular}{lll}
        & Figure \ref{fig:katazynski_2006_fig3} & Figure \ref{fig:katazynski_2006_fig4} \\
        \hline\hline
        $\delta$ & 21 & 33 \\
        B & 0.05 & 0.11 \\
        R & $3.2 \times 10^{15}$ & $10^{15}$ \\
        $n_0$ & 7 & $5\times 10^{-2}$ \\
        $\gamma_1$ & 1 & 1 \\
        $\gamma_2$ & 2 & 2 \\
        $Q_{inj}$ & 0 & $5 \times 10^{-2}$ \\
        $t_{esc}$ & $\infty$ & $t_{acc}$ \\
    \end{tabular}
    \caption{Parameters used for Figs. \ref{fig:katazynski_2006_fig3} and \ref{fig:katazynski_2006_fig4} }
    \label{tab:steady_state_2_params}
\end{table}

\begin{figure*}
    \centering
    \includegraphics[width=0.32\linewidth]{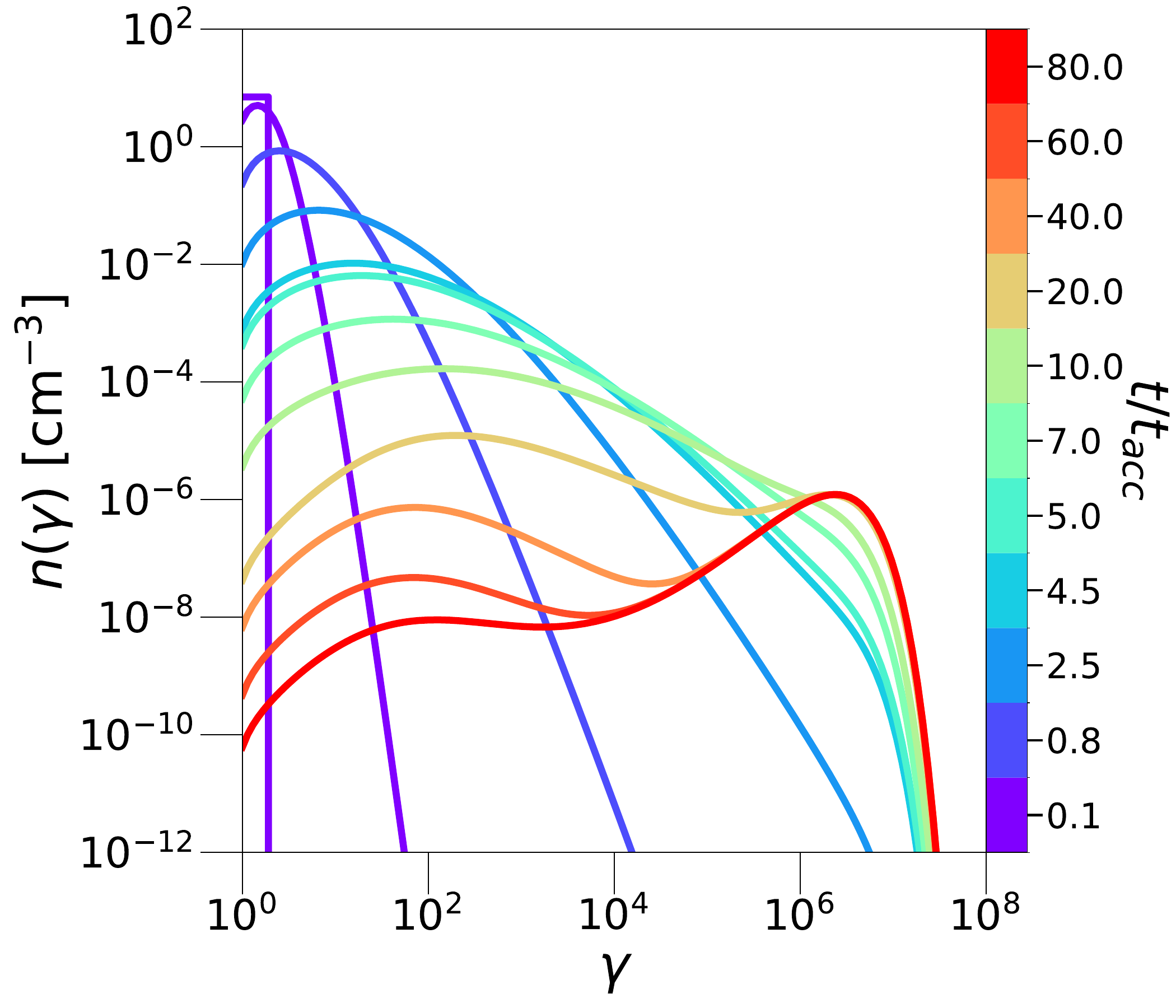}
    \includegraphics[width=0.32\linewidth]{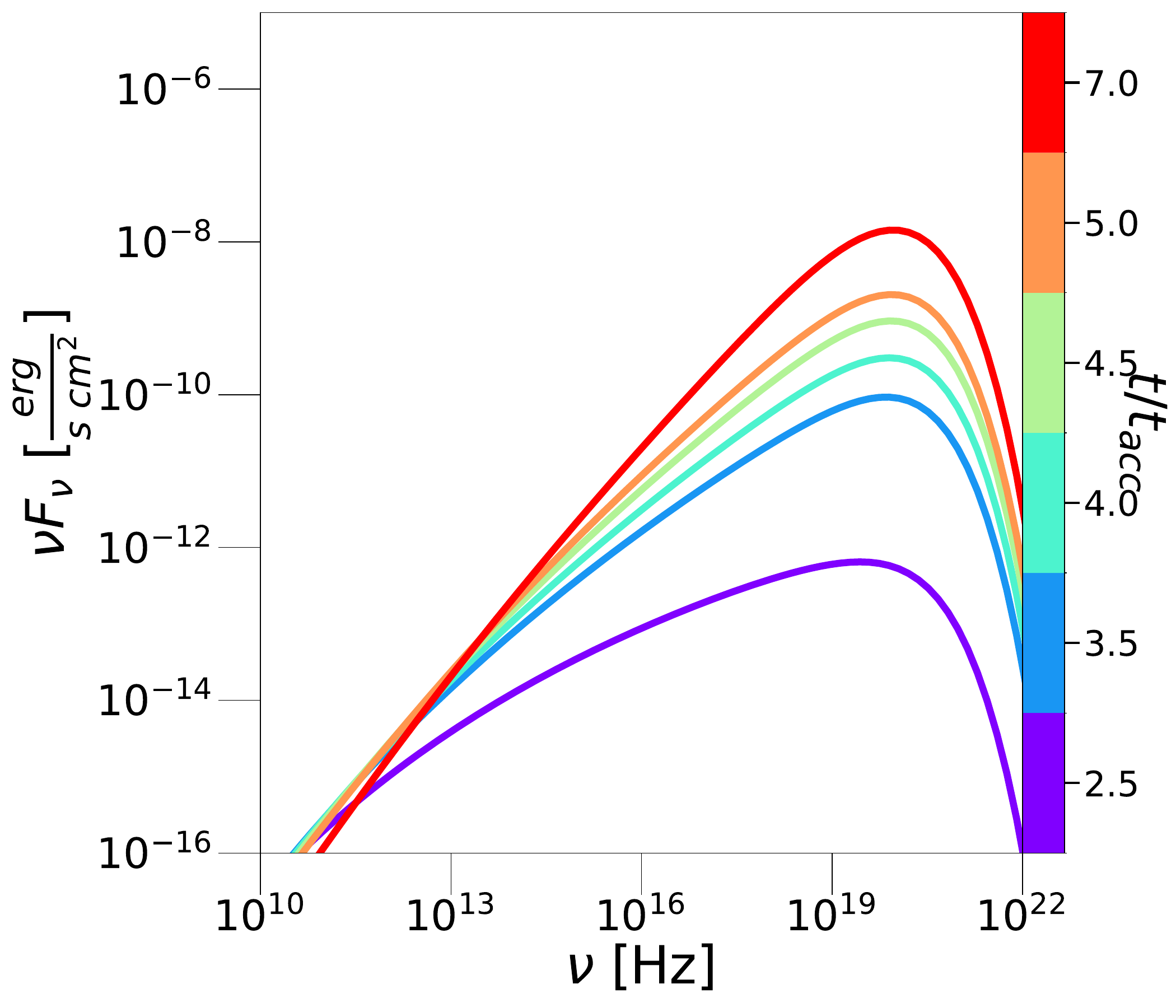}
    \includegraphics[width=0.32\linewidth]{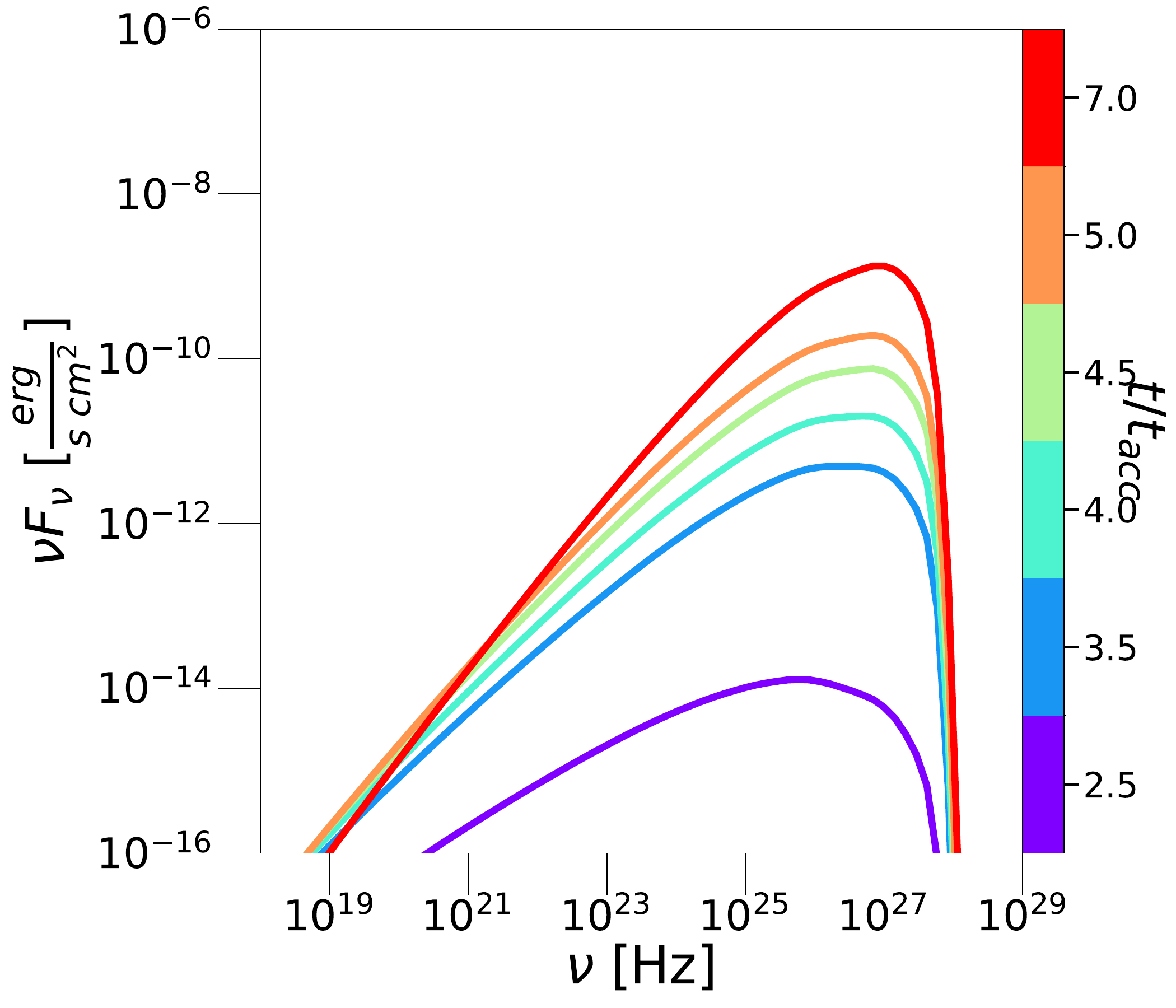}
    \caption{As closely as possible, we recreate Figure 3 from \citet{Katarzynski:2006gh}, with parameters summarized in Table \ref{tab:steady_state_2_params}. \emph{Left}: The particle distribution does not experience escape or injection, but the distribution remains complex due to SSC cooling being more efficient at lower energies, producing a double peak distribution. \emph{Middle}: Corresponding synchrotron emission. \emph{Right}: Resultant self-synchrotron Compton emission.}
    \label{fig:katazynski_2006_fig3}
\end{figure*}

\begin{figure*}
    \centering
    \includegraphics[width=0.32\linewidth]{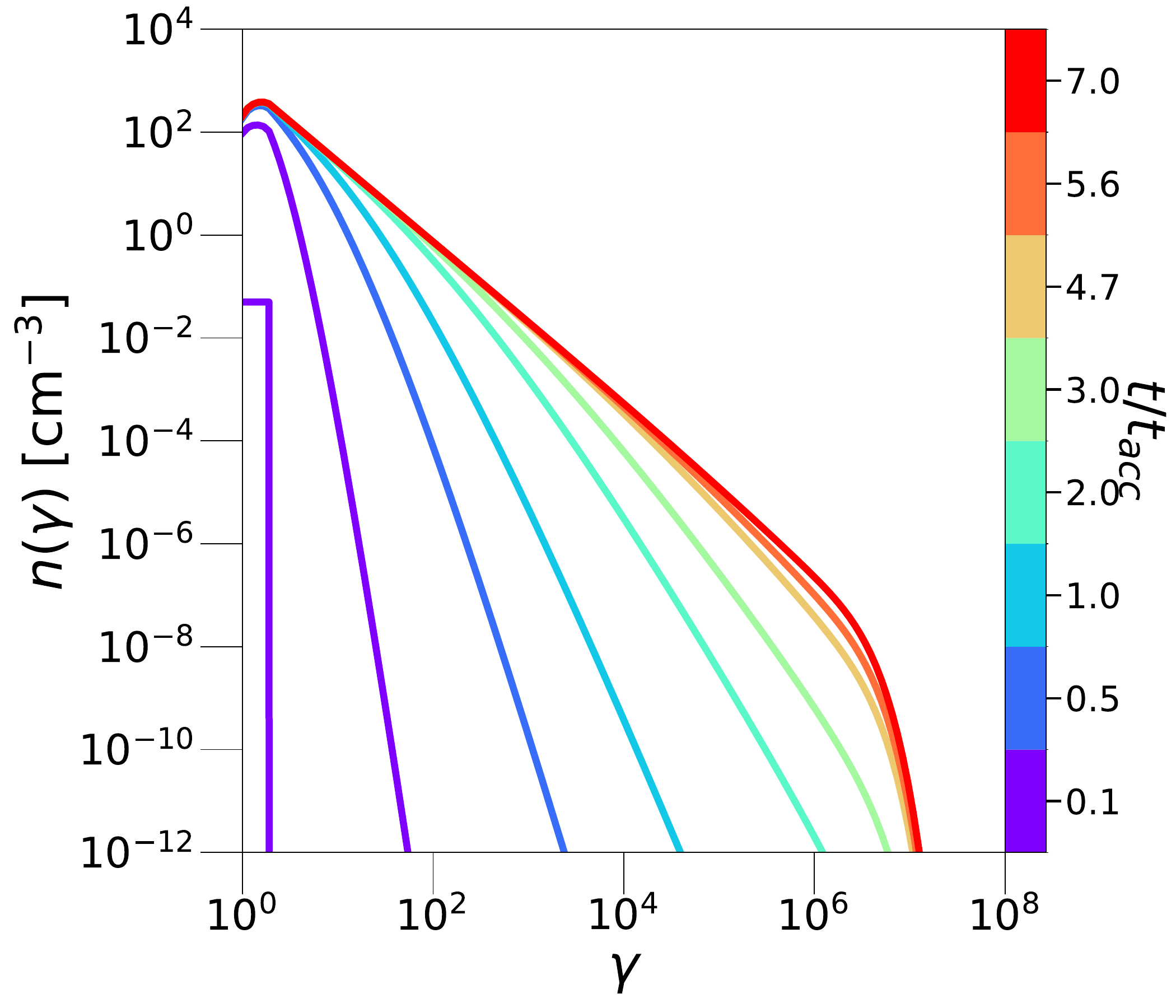}
    \includegraphics[width=0.32\linewidth]{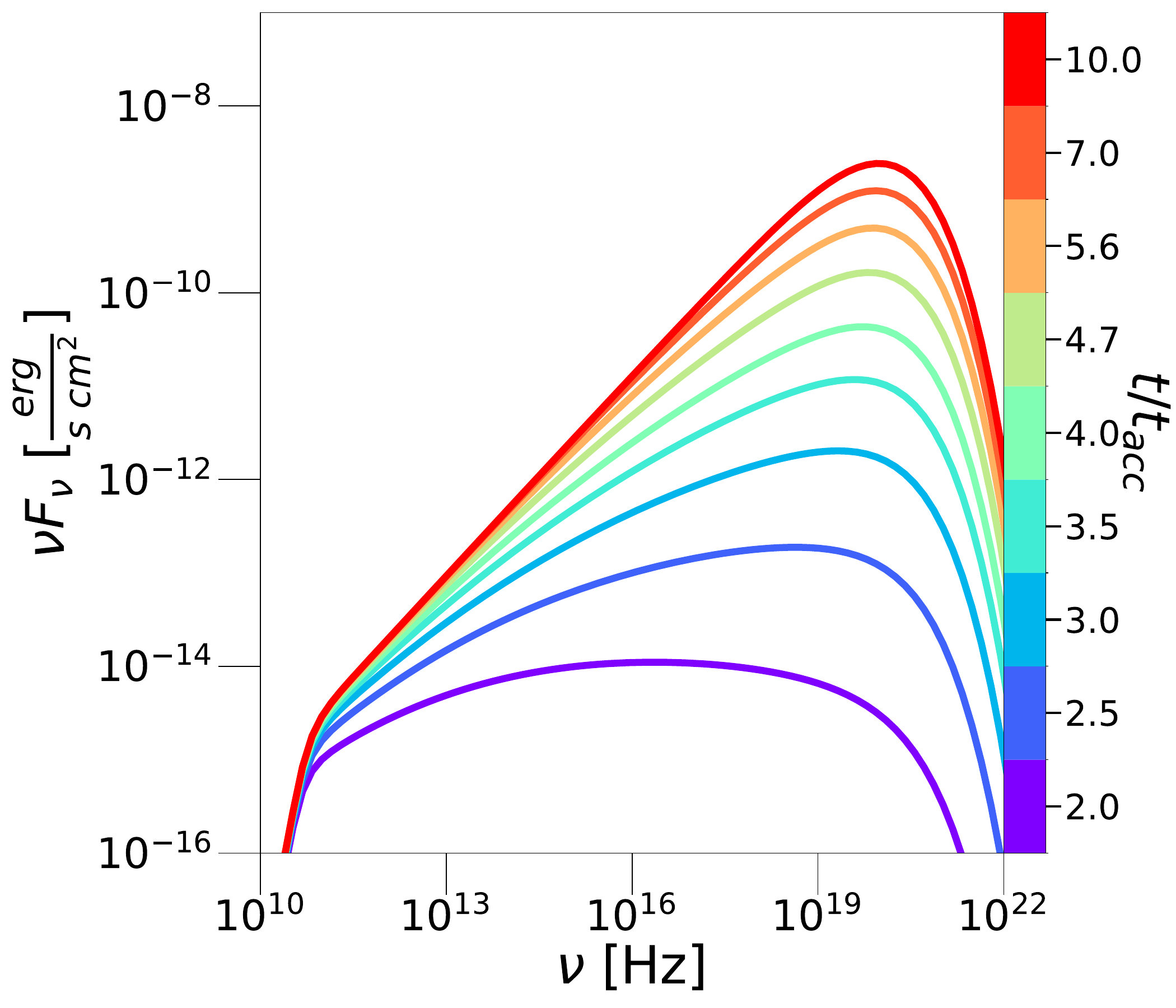}    \includegraphics[width=0.32\linewidth]{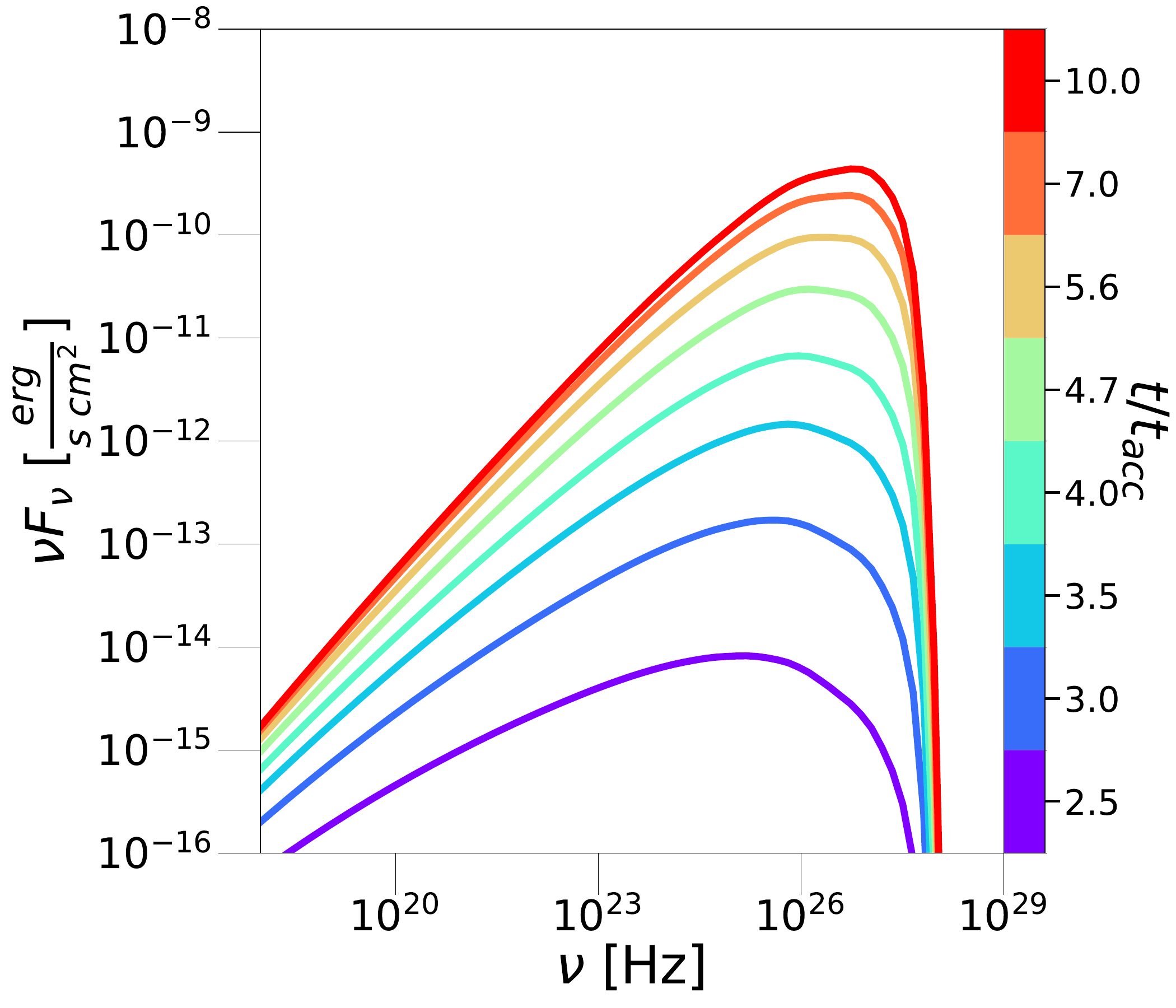}
    \caption{As closely as possible, Figure 4 from \citet{Katarzynski:2006gh} is recreated, with parameters summarized in Table \ref{tab:steady_state_2_params}. \emph{Left}: Particles do not experience escape or injection, but the distribution remains complex due to SSC cooling being more efficient at lower energies, producing a double peak distribution. \emph{Middle}: Corresponding synchrotron emission. \emph{Right} Resultant self-synchrotron Compton emission.}
  \label{fig:katazynski_2006_fig4}
\end{figure*}

The results in Figs. \ref{fig:katazynski_2006_fig3} and \ref{fig:katazynski_2006_fig4} align well with the results presented in \citet{Katarzynski:2006gh}, but a direct replication is impossible without specific details about the radiation transfer used and their model for extragalactic background light (EBL) absorption. Regardless, even with these unknowns and switching to full KN cooling, a qualitative agreement is achieved. In the top of Fig. \ref{fig:katazynski_2006_fig3}, we observe, as in \citet{Katarzynski:2006gh}, that the particle distribution can become complex when introducing SSC cooling. The double peak observed in the top of Fig. \ref{fig:katazynski_2006_fig3} is a result of the SSC cooling being dominant at lower energies, while at higher energies, synchrotron cooling prevails. When injection and escape are considered (Fig. \ref{fig:katazynski_2006_fig4}, left panel), the complex behavior from SSC cooling is obscured, and the distribution appears more like the case shown in top of Fig. \ref{fig:katazynski_2006_fig2}.

\section{Conclusion}
\label{sec:conclusion}

In this work, we have introduced an open-source particle evolution and radiation solver called \tleco. We began by discussing the Chang-Cooper schema \citep{Chang:1970co} used for evaluating particle evolution and assessed numerical accuracy through convergence testing against analytical solutions. Subsequently, we explored various radiative tools included in the software suite, focusing particularly on synchrotron and inverse Compton cooling and emission algorithms. Finally, we demonstrated the capabilities of \tleco\ by reproducing results from \citet{Katarzynski:2006gh}.

In modern high-energy astrophysics, where large surveys yield numerous discoveries, rapid prototyping of theories for comparison with observations is essential. As a Python library, \tleco\ leverages the flexibility and readability of Python, along with the extensive collection of available Python libraries. This combination reduces development time when constructing numerical models with \tleco, without sacrificing the advantages of compiled languages, thanks to the underlying Rust code. \tleco\ serves as a toolkit to aid researchers in building their own simulations, which can include physics not currently implemented in \tleco. Although additional physics can be easily integrated on an ad hoc basis, the ultimate goal is to continue expanding \tleco. A list of planned future developments can be found in the GitHub repository \href{https://github.com/zkdavis/Tleco}{https://github.com/zkdavis/Tleco}.



\begin{acknowledgments}
ZD and DG acknowledge support from the NSF AST-2107802, AST-2107806 and AST-2308090 grants.
\end{acknowledgments}

%



\software{astropy \citep{Astropy2022},
          numpy \citep{numpy2020array},
          }





\bibliography{Ppaper}
\bibliographystyle{aasjournal}



\end{document}